\documentclass[11pt,letterpaper]{article}
\usepackage[margin=1in]{geometry}
\usepackage{amsmath,amssymb}
\usepackage{bm}
\usepackage{graphicx}
\usepackage{booktabs}
\usepackage{multirow}
\usepackage{placeins}
\usepackage{float}
\usepackage{cite}
\usepackage[colorlinks=true,linkcolor=blue,citecolor=blue,urlcolor=blue]{hyperref}
\hypersetup{%
	pdfauthor={Shirin Hosseinmardi, Xiangyu Sun, Ramin Bostanabad},
	pdftitle={Linear Elasticity Versus Finite-Strain Hencky Models in Multi-Material Thermo-Mechanical Topology Optimization},
	pdfkeywords={topology optimization, thermo-mechanical design, geometric nonlinearity, Hencky strain, temperature-dependent properties, compliant mechanisms},
	pdfsubject={Comparative study of constitutive and property-model fidelity in thermo-mechanical topology optimization},
}

\newcommand{\keywords}[1]{\par\smallskip\noindent \small \textbf{Keywords:} #1\par}

\newcommand{\uout}{u_{\mathrm{out}}}
\newcommand{\uref}{u_{\mathrm{out}}^{\mathrm{ref}}}
\newcommand{\TD}{T_{D}}
\newcommand{\Tinf}{T_{\infty}}
\newcommand{\epsth}{\varepsilon_{\mathrm{th}}}
\newcommand{\dlaw}{\Delta_{\mathrm{law}}}
\newcommand{\dprop}{\Delta_{\mathrm{prop}}}

\title{On the Importance of Geometric Nonlinearity and Temperature-Dependent Properties in Multi-Material Thermo-Mechanical Topology Optimization}

\author{%
  Shirin Hosseinmardi$^{1}$ \quad Xiangyu Sun$^{1}$ \quad Ramin Bostanabad$^{1,2,\ast}$\\[6pt]
  \small $^{1}$Department of Mechanical and Aerospace Engineering, University of California,  Irvine\\
  \small $^{2}$Department of Civil and Environmental Engineering, University of California, Irvine\\[4pt]
  \small $^{\ast}$Corresponding author \texttt{(raminb@uci.edu)}
}
\date{}

\begin{document}

\maketitle

\begin{abstract}
Thermo-mechanical compliant devices are commonly designed with small-strain linear elasticity and temperature-independent material properties, even though they might operate hundreds of kelvin above ambient where both assumptions are questionable. In this work, we quantify the effect and cost of each assumption in multi-material topology optimization of thermally actuated compliant devices. To this end, we introduce a physics-informed, simultaneous analysis-and-design framework with (i) a finite-strain quadratic-Hencky (logarithmic-strain) constitutive model whose isotropic thermal eigenstrain admits an exact additive split in log-strain space, and (ii) temperature-dependent conductivity, thermal expansion, and elastic moduli for a titanium--copper--steel material system. We optimize a thermal actuator and a thermal gripper at three design temperatures under both a baseline model and the full physics, subject to mass and manufacturability constraints. Every converged design is re-evaluated by verified nonlinear finite element solvers in the full factorial of constitutive law and property model. The comparison between the two factors reveals that the constitutive law is the decisive modeling choice: These devices work as linkages where linear kinematics mistakes rotation for compressive strain; its error therefore grows with the design temperature and concentrates on the very layouts that exploit rotation best. Because a linear optimizer also steers away from the rotation-rich mechanisms that would expose this bias, the model can deceptively appear trustworthy when validated against its own designs. Designing with the full physics yields consistently stronger and more temperature-robust devices at a modest increase in design-time cost.\footnote{Our data, models, and codes are accessible via our \href{https://github.com/Bostanabad-Research-Group/m-PIGP-thermomech}{GitHub repository}.}
\end{abstract}


\section{Introduction}
Topology optimization (TO) is an established computational design method that finds the optimal spatial distribution of material in a design domain directly from the governing physics and design requirements. TO has matured into a standard tool for designing compliant mechanisms, thermal devices, and multi-physics systems \cite{bendsoe1988, bendsoe2013, sigmund2013, sigmund1997, deaton2014}. In thermo-mechanical TO, actuation is generated by constrained thermal expansion rather than external loads, a principle widely exploited in micro-electromechanical systems (MEMS) where electrothermal actuators and grippers convert Joule heating or a prescribed thermal environment into controlled strokes \cite{jonsmann1999, sigmund2001actuator, xia2018, du2009}. The design space is considerably richer in multi-material settings, where the optimizer arranges several candidate metals to exploit their unique material properties subject to mass and manufacturability constraints \cite{zuo2017, sun2026}.

Nearly all relevant literature, including ours \cite{sun2026}, rests on two modeling conveniences. First, material properties are held constant and second the mechanical response is described by small-strain linear elasticity, even though the slender members and distributed hinges that make a compliant mechanism effective undergo moderate rotations well before strains become large \cite{buhl2000, pedersen2001, bruns2001}.

\textbf{Material Properties.} In the thermal-actuator literature the constant values are often nondimensionalized and independent of the operating temperature \cite{sigmund2001actuator,
jonsmann1999, yin2002, xia2018}. Investigations that do use physical properties, adopt \emph{room temperature} values despite the mismatch with operating temperatures~\cite{gao2016topology,deaton2016}. Deng and
Suresh~\cite{deng2017}, for example, state that their reference room temperature serves ``only'' to select material properties, and then
optimize a titanium-alloy wing rib at $\Delta T = 270$ K; our own prior
work~\cite{sun2026} likewise uses room-temperature moduli and
conductivities at a $673$ K boundary temperature. The convention is
particularly damaging: at $673$ K, copper (the dominant phase in the optimized devices of
this work) has an elastic modulus $15\%$ below its room-temperature value and a mean
coefficient of thermal expansion $9\%$ above it, while across the candidate metals the
moduli drop by $15\text{--}21\%$ and the thermal conductivity of steel falls by $26\%$
\cite{ho1972, touloukian1975, chang1966, dever1972}. A room-temperature-anchored model
misrepresents not only the stiffness of the device but also the thermal load that drives it.

Although restricted to linear elasticity, a small body of work does use temperature-dependent properties and reports
concrete gains. Tang et al.~\cite{tang2023} make conductivity, elastic tensor, and expansion coefficient temperature-dependent under large gradients, carrying the dependence through the adjoint, and find constant-property optimization significantly inaccurate; Zheng et al.~\cite{zheng2023} reach the same
conclusion over $293$--$773$ K. For multi-material design, Chen et
al.~\cite{chen2022} report that the temperature-dependent modulus dominates compliance while the expansion coefficient principally reshapes the material distribution compared against room-temperature-anchored baselines.
%==============================================================
\begin{figure*}[h]
\centering
\includegraphics[width=\textwidth]{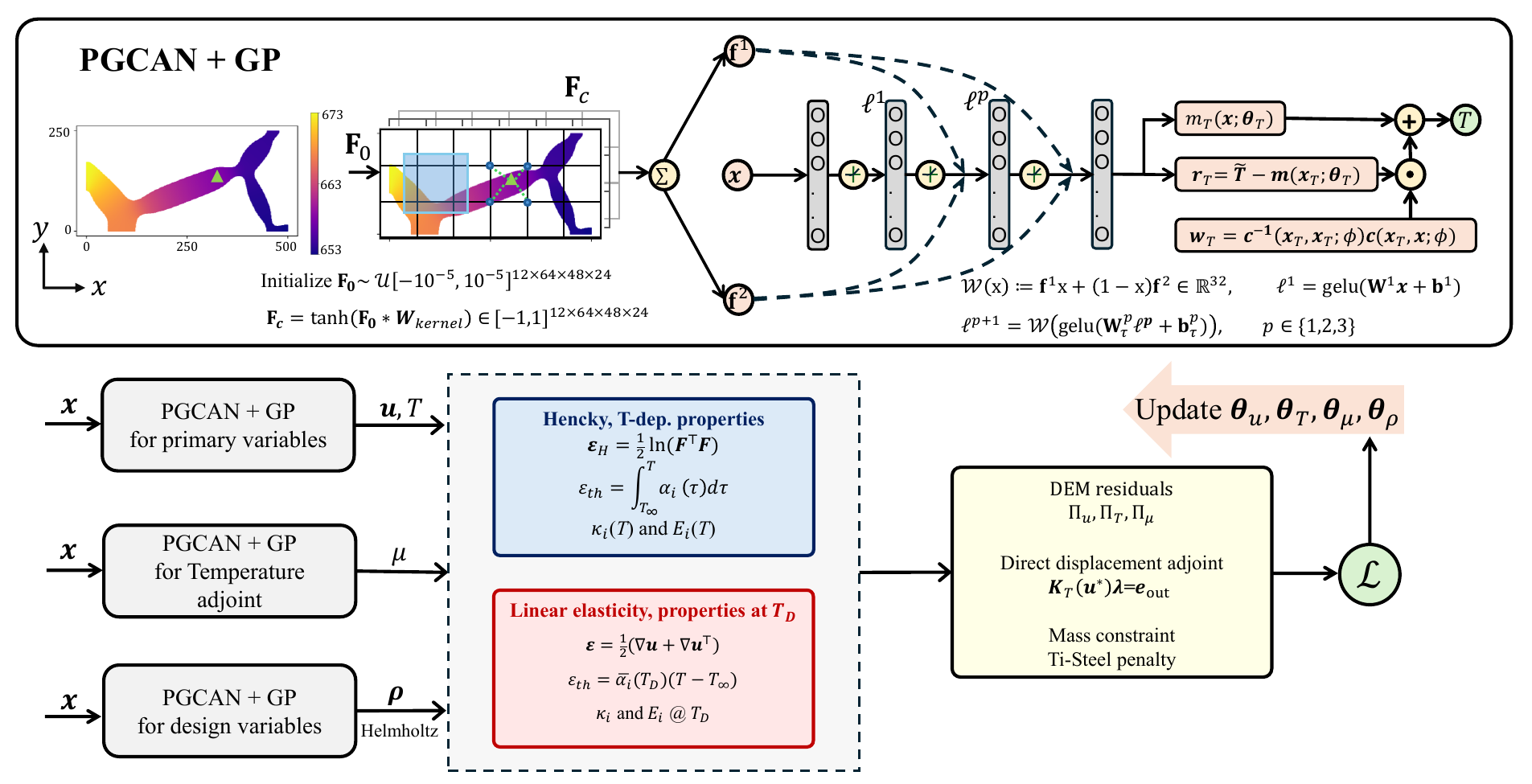}
\caption{Overview of m-PIGP framework: PGCAN-parameterized GP priors represent the primal fields ($\bm{u}$, $T$), the temperature adjoint $\mu$, and the design field $\boldsymbol{\rho}$ (pushed through the Helmholtz filter); The PGCAN encoder convolves the trainable feature tensor $\mathbf{F}_0$ to obtain  $\mathbf{F}_c$, then interpolates features surrounding a query point via cosine interpolation, and passes them to 
a shallow attention-modulated decoder. Gaussian process conditioning enforces Dirichlet BCs on the state and adjoint fields and prescribes phases where required. The outputs enter one of the two physics models compared in this work: the baseline branch (linear kinematics with properties anchored at $\TD$) or the high-fidelity branch (quadratic-Hencky kinematics with temperature-dependent properties) whose deep-energy residuals are combined with the direct displacement adjoint of Eq.~\eqref{eq:uadj}, the mass constraint, and the Ti--Steel interface penalty into the training loss $\mathcal{L}$ of Eq.~\eqref{eq:loss}, minimized simultaneously over all parameters with Adam.}
\label{fig:flowchart}
\end{figure*}
%==============================================================
A better constant-property practice evaluates the properties at the design temperature; we therefore adopt this stronger convention as the baseline of this study and quantify the potential benefit of incorporating fully temperature-dependent properties. Both modeling conveniences bias the very objective that the optimizer maximizes, i.e., the predicted output stroke. Since the optimizer is free to exploit any inaccuracy of its physics model, such biases do not merely misreport performance; they can also steer the search toward designs whose advantage is an artifact of the model. Quantifying these effects, separately and jointly, is the main goal of this paper.

\textbf{Constitutive Model.}
Geometrically nonlinear TO of compliant mechanisms has an established history built on Green--Lagrange kinematics with St.~Venant--Kirchhoff or neo-Hookean energies \cite{buhl2000, pedersen2001, bruns2001}, together with interpolation schemes that stabilize low-density regions at finite strain \cite{wang2014}. Extensions to \emph{thermally} driven finite-strain problems are comparatively recent: Chung et al.\ \cite{chung2020} combined level-set TO with nonlinear thermoelasticity, Sui et al.\ \cite{sui2023} designed thermo-hyperelastic structures via inverse-motion form finding, and Granlund et al.\ \cite{granlund2024} optimized compliant mechanisms under transient thermal loading with a multi-material neo-Hookean model. Temperature-dependent properties have likewise received attention in thermo-elastic TO under large temperature gradients \cite{tang2023}. The Hencky (logarithmic) strain measure, however, has essentially not been used in thermo-mechanical TO; its appearances in TO are limited to purely mechanical settings such as hyperelastic ground structures \cite{ramos2015} and material-point-method TO of large-deformation mechanisms \cite{padhy2026}. This measure is particularly well suited to thermo-mechanical design: for isotropic thermal expansion, the logarithmic thermal eigenstrain acts as an exact additive and isotropic offset to the elastic strain across arbitrary deformations, unlike the traditional Green–Lagrange eigenstrain, which is only first-order accurate with respect to thermal stretch. The quadratic-Hencky energy is furthermore an excellent model of real metals up to moderate elastic strains \cite{hencky1924, anand1979, neff2016}, precisely the regime of metallic compliant devices.

\textbf{TO via Machine Learning.} The majority of TO frameworks for thermomechanical design are formulated with classic density-based (SIMP) \cite{bendsoe1988, sigmund2013} or level-set \cite{xia2018} methods and are implemented in a nested analysis and design fashion: every design update requires solving the governing PDEs to convergence, sensitivities follow from a separately derived adjoint problem and the design advances through its own update loop \cite{bendsoe2013,deaton2014}. More recently, machine learning (ML) techniques have emerged as promising tools for TO. Early efforts focused primarily on data-driven surrogates or generative models trained on datasets produced by conventional TO solvers \cite{kallioras2020,nie2021}. Although these approaches can reduce computational cost, they remain tied to the physics represented in their training data, which makes them ill-suited to interrogating the physics. Physics-informed ML (PIML) offers a more flexible paradigm in which the governing equations and the design objective are embedded directly into the learning process \cite{raissi2019,mora2025}. PIML provides mesh-free representations of the state and design fields that stay consistent with the underlying physics, accommodates additional design constraints as auxiliary loss terms and is differentiable end-to-end. Consequently, the constitutive model reduces to an energy density in the
loss, so upgrading the physics means modifying one term rather than a solver overhaul.

Building on this line of work, we have developed simultaneous, mesh-free TO frameworks based on physics-informed Gaussian processes (PIGPs) whose mean functions are parameterized with specialized neural networks \cite{yousefpour2025, sun2025, mora2025, shishehbor2024} and extended them to multi-material and multi-physics problems \cite{sun2026}. 

\textbf{Contributions.} All our previous works adopt linear kinematics and constant properties, and none enforces manufacturability. Herein we close both gaps: it upgrades and interrogates the physics, and it does so under the manufacturing constraints (i.e., a minimum feature size and the exclusion of metallurgically incompatible material interfaces) that favor realizable multi-material thermo-mechanical designs.
In summary, our unique contributions are:
\begin{itemize}
\item Building an ML-based TO framework considering finite deformations, multi-material designs, and manufacturability constraints, see Fig.~\ref{fig:flowchart}.
\item Introducing per-phase temperature-dependent conductivity, expansion, and modulus for a \{Ti, Cu, Steel\} system, fit to published measurements over $293\text{--}1100$ K \cite{ho1972, touloukian1975, fisher1964, chang1966, dever1972, incropera2007}, with the eigenstrain computed as the exact integral of the instantaneous coefficient and every property--temperature pathway carried into the adjoint.
\item Designing a controlled comparison between the two physics models and quantifying both modeling errors with temperature-robustness and transferability analyses: Anchoring the constant-property baseline at the design temperature isolates the constitutive law as the decisive modeling choice. 
\item Enforcing manufacturability throughout: a Helmholtz-type filter \cite{lazarov2011} for minimum feature size and an interface-exclusion penalty preventing brittle Ti--Fe intermetallics \cite{moshokoa2024} so that the physics comparison is conducted over manufacturable layouts; we refer to the resulting framework as m-PIGP.
\end{itemize}

The remainder of the paper is organized as follows. Section~\ref{sec:method} presents the governing physics, the two constitutive models, the temperature-dependent property model, and the optimization framework. Section~\ref{sec:results} describes the cross-evaluation protocol and reports the optimized designs and the comparative studies. Section~\ref{sec:conclusion} summarizes the findings and outlines future work.

\FloatBarrier
\section{Methodology}\label{sec:method}
We consider multi-material thermo-mechanical TO in a 2D domain $\Omega \subset \mathbb{R}^2$ occupied by $n_m$ candidate materials and void. The design is the vector field $\boldsymbol{\rho}(\bm{x}) = [\rho_0, \rho_1, \ldots, \rho_{n_m}]^\top$ of local volume fractions with $\sum_{i=0}^{n_m} \rho_i(\bm{x}) = 1$, where $i=0$ denotes void. Phase-dependent properties are interpolated with the power law
\begin{equation}\label{eq:simp}
\mathcal{P}(\bm{x}, T) = \sum_{i=0}^{n_m} \mathcal{P}_i(T)\, \rho_i^{\,p}(\bm{x}),
\end{equation}
where $\mathcal{P} \in \{E, \kappa, q_v\}$ collects the Young's modulus, thermal conductivity, and volumetric heat sink, $p$ is the penalization exponent (continuated from 1 to 3), and the coefficient of thermal expansion is interpolated linearly (i.e., no penalty) to avoid biasing the eigenstrain of intermediate densities. We use a uniform Poisson's ratio of $\nu = 0.31$ and temperature-dependent phase values $\mathcal{P}_i(T)$ as detailed in Sec.~\ref{sec:tdep}.

\FloatBarrier
\subsection{Governing Thermo-Mechanical Physics}\label{sec:physics}

\begin{figure*}[h]
\centering
\includegraphics[width=\textwidth]{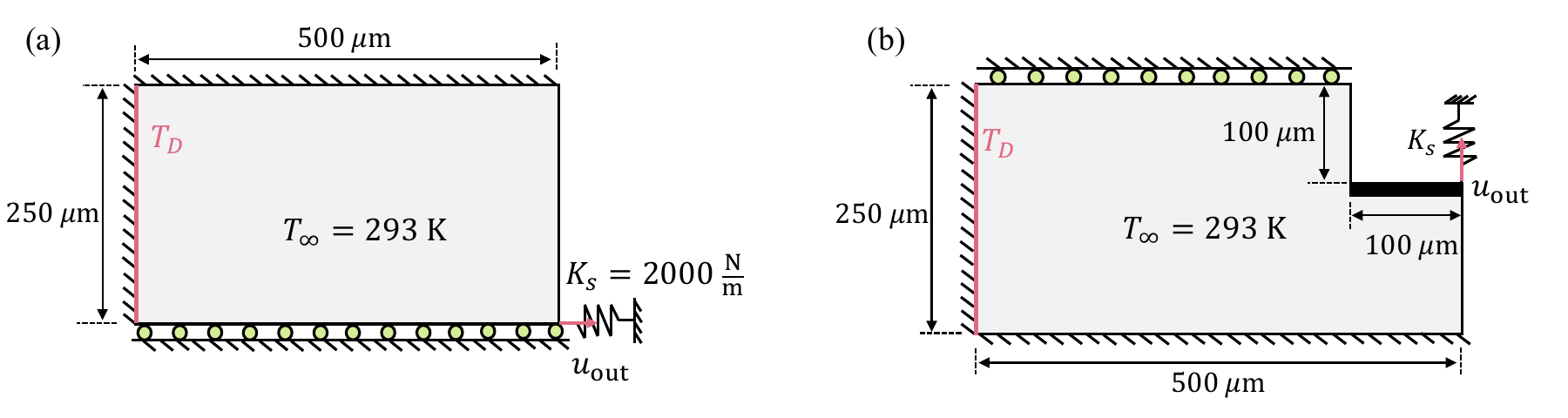}
\caption{Design domains and boundary conditions for (a) the thermal actuator and (b) the thermal gripper. Both are half layouts, symmetric about the roller edge. Actuation is driven by the temperature $\TD$ prescribed on the left edge; all other boundaries are adiabatic and a volumetric heat sink $q_v$ acts in the solid. The output port is restrained by a spring $K_s$ and the objective is maximizing the output stroke $\uout$. The layout thickness is $15\,\mu$m.}
\label{fig:domain}
\end{figure*}

Figure~\ref{fig:domain} shows the two benchmark devices together with their boundary conditions (BCs). Each occupies a $500 \times 250\,\mu$m half domain of thickness $15\,\mu$m; the gripper additionally removes a $100\times100\,\mu$m notch that forms the jaw. Heat transfer is by steady conduction,
\begin{equation}\label{eq:heat}
-\nabla \cdot \big[\kappa(\bm{x}, T)\, \nabla T\big] = q_v(\bm{x}), \quad \forall\, \bm{x} \in \Omega,
\end{equation}
with $T = \TD$ prescribed on the left edge, adiabatic conditions elsewhere, and a distributed volumetric heat sink $q_v = -4.5\times10^{-8}\,\mathrm{W}\,\mu\mathrm{m}^{-3}$ in the solid phases that lumps losses to the surroundings. Since $\kappa$ depends on $T$, Eq.~\eqref{eq:heat} is mildly nonlinear and is treated by Picard iteration: properties are evaluated at the current (frozen) temperature so that the fixed point satisfies the nonlinear weak form exactly. In the simultaneous solver of Sec.~\ref{sec:framework}, the temperature field is trained on the deep-energy functional
\begin{equation}\label{eq:PiT}
\Pi_T = \int_\Omega \tfrac{1}{2}\, \kappa(\bm{x}, T^*)\, |\nabla T|^2\, \mathrm{d}V - \int_\Omega q_v\, T\, \mathrm{d}V,
\end{equation}
whose stationarity at the Picard fixed point $T^* = T$ recovers the weak form of Eq.~\eqref{eq:heat}.
%==============================================================
%==============================================================
Mechanical equilibrium is written in total-Lagrangian form. With displacement $\bm{u}(\bm{x})$, displacement gradient $\mathbf{H} = \nabla \bm{u}$, and deformation gradient $\mathbf{F} = \mathbf{I} + \mathbf{H}$, equilibrium follows from stationarity of the potential energy
\begin{equation}\label{eq:potential}
\Pi_u = \int_\Omega \psi\big(\mathbf{H}; \boldsymbol{\rho}, T\big)\, \mathrm{d}V + \tfrac{1}{2} K_s \big(\bm{u}(\bm{x}_{\mathrm{out}}) \cdot \bm{e}_s\big)^2,
\end{equation}
where $\psi$ is the strain-energy density of the chosen constitutive model, $K_s = 2000$ N/m is the workpiece spring at the output port $\bm{x}_{\mathrm{out}}$, and $\bm{e}_s$ is the output direction. No external tractions act; deformation is driven entirely by the thermal eigenstrain carried inside $\psi$. The design objective is to maximize the output stroke $\uout = \bm{u}(\bm{x}_{\mathrm{out}}) \cdot \bm{e}_s$.

\FloatBarrier
\subsection{Constitutive Models: Linear Versus Quadratic Hencky}\label{sec:constitutive}
The two constitutive models compared in this work share the same plane-stress elastic coefficients $C_1 = E/(1-\nu^2)$ and $C_2 = E/[2(1+\nu)]$ and the same isotropic thermal eigenstrain $\epsth$, and differ only in kinematics.

\textbf{Linear elasticity (small strain).} The strain is $\boldsymbol{\varepsilon} = \tfrac{1}{2}(\mathbf{H} + \mathbf{H}^\top)$ and the energy density is the standard quadratic form
\begin{equation}\label{eq:psiL}
\psi_L(\mathbf{H}) = \tfrac{1}{2}\, \big(\boldsymbol{\varepsilon} - \epsth \mathbf{I}\big) : \mathbb{C} : \big(\boldsymbol{\varepsilon} - \epsth \mathbf{I}\big),
\end{equation}
with $\mathbb{C}$ the plane-stress stiffness. This is the kinematic model used in most of the thermo-mechanical TO literature.

\textbf{Quadratic Hencky (finite strain).} The material logarithmic strain is $\boldsymbol{\varepsilon}_H = \tfrac{1}{2} \ln (\mathbf{F}^\top \mathbf{F})$, and the energy is the same quadratic form evaluated on the log strain,
\begin{equation}\label{eq:psiH}
\psi_H(\mathbf{H}) = \tfrac{1}{2}\, \big(\boldsymbol{\varepsilon}_H - \epsth \mathbf{I}\big) : \mathbb{C} : \big(\boldsymbol{\varepsilon}_H - \epsth \mathbf{I}\big).
\end{equation}
The quadratic-Hencky model is frame indifferent, captures finite rotations exactly, and reproduces measured metal response up to moderate elastic strains \cite{anand1979, neff2016}; it reduces to Eq.~\eqref{eq:psiL} as $\mathbf{H} \to \mathbf{0}$. Two features make it particularly attractive for thermo-mechanical design. First, for isotropic expansion the multiplicative thermal split $\mathbf{F} = \mathbf{F}_e(\vartheta \mathbf{I})$ with thermal stretch $\vartheta = \exp\!\big(\int_{\Tinf}^{T} \alpha\, \mathrm{d}\tau\big)$ commutes, so
\begin{equation}\label{eq:split}
\ln \mathbf{U} = \ln \mathbf{U}_e + \epsth\, \mathbf{I}, \qquad \epsth = \int_{\Tinf}^{T} \alpha(\tau)\, \mathrm{d}\tau,
\end{equation}
i.e., the elastic log strain is obtained from the total log strain by subtracting an additive isotropic eigenstrain \emph{exactly} at finite deformation, the property that makes Eqs.~\eqref{eq:psiL} and \eqref{eq:psiH} formally identical and isolates the strain measure as the only difference between the two models. Second, in two dimensions $\boldsymbol{\varepsilon}_H$ admits a closed form: writing $\mathbf{C} = \mathbf{F}^\top\mathbf{F}$ with invariants $I_1 = \operatorname{tr}\mathbf{C}$ and $\det \mathbf{C}$, we use
\begin{equation}\label{eq:closedform}
\boldsymbol{\varepsilon}_H = \tfrac{1}{4} \ln (\det \mathbf{C})\, \mathbf{I} + \frac{f(q)}{I_1} \Big(\mathbf{C} - \tfrac{1}{2} I_1 \mathbf{I}\Big),
\end{equation}
where $q = \big[(C_{11}{-}C_{22})^2 + 4C_{12}^2\big]/I_1^2$ and $f(q) = \operatorname{atanh}(\sqrt{q})/\sqrt{q}$ is smooth on $[0,1)$. This parameterization avoids eigendecomposition, remains finite and differentiable at coalescent stretches ($\mathbf{C} \propto \mathbf{I}$) and admits stable second derivatives, thereby removing the matrix-logarithm differentiation cost that has discouraged Hencky-strain adjoints in TO \cite{padhy2026}.

Because void regions cannot be allowed to distort or invert at finite strain, our element energy uses the three-term interpolation of Wang et al.\ \cite{wang2014},
\begin{equation}\label{eq:wang}
\psi = \psi_H(\gamma \mathbf{H}) + \psi_L(\mathbf{H}) - \psi_L(\gamma \mathbf{H}),
\end{equation}
where $\gamma(\bm{x}) \in [0,1]$ is a sharp smooth-Heaviside indicator of the local solid fraction. Solid material ($\gamma \approx 1$) sees the full nonlinear energy while void ($\gamma \approx 0$) sees small-strain elasticity with the void stiffness. 

We emphasize that the popular two-term shortcut $\psi_H(\gamma\mathbf{H}) + (1{-}\gamma^2)\psi_L(\mathbf{H})$ relies on the homogeneity $\psi_L(\gamma\mathbf{H}) = \gamma^2 \psi_L(\mathbf{H})$, which \emph{fails} in the presence of a thermal eigenstrain (the cross term is linear in strain); the three-term form of Eq.~\eqref{eq:wang} is exact for any $\epsth$ and cancels the spurious pure-eigenstrain energy at $\gamma = 0$ identically. Under the linear model, Eq.~\eqref{eq:wang} collapses to $\psi_L(\mathbf{H})$ and $\gamma$ is inert.

\FloatBarrier
\subsection{Temperature-Dependent Material Properties}\label{sec:tdep}

Our candidate set is \{void, Ti, Cu, Steel\}, three metals that span a broad range of conductivity, expansion, and stiffness; giving the optimizer genuine material contrast to exploit. All three materials remain solid over the design temperatures considered here: at the highest, $\TD = 1073$ K, copper is the most severely loaded at ${\sim}0.8$ of its melting temperature ($1358$ K), while steel ($\approx 1800$ K) and titanium ($1941$ K) sit near $0.6$ and $0.55$ of theirs. 

Titanium stays in its $\alpha$ phase throughout, its transus lying at $1155$ K, so a single set of property fits covers the whole window. At such homologous temperatures the metals would creep under sustained load, so the devices are understood to operate intermittently, in short actuation cycles as is typical of thermally driven MEMS. This is the regime in which the rate-independent elastic response of Eqs.~\eqref{eq:psiL} and \eqref{eq:psiH} applies, and it is the operating mode assumed throughout this work. 

The room-temperature base values are $\kappa = 21.9$, $400$, and $60$ W/(m$\,$K), $\alpha = 8.6$, $17$, and $12 \times 10^{-6}$ K$^{-1}$, and $E = 115$, $128$, and $200$ GPa for Ti, Cu, and the steel, respectively, with mass densities $\bar{\rho} = 4.51$, $8.96$, and $7.80$ g/cm$^3$ \cite{ho1972, incropera2007, touloukian1975, fisher1964, chang1966, dever1972}. The third phase is a plain carbon steel, and is labeled Steel throughout. Each phase property is written as its room-temperature base value multiplied by a polynomial correction factor in $\Delta T = T - 293$ K,
\begin{equation}\label{eq:propfactor}
\begin{split}
\kappa_i(T) &= \kappa_i\, f_{\kappa,i}(\Delta T), \quad
\alpha_i(T) = \alpha_i\, f_{\alpha,i}(\Delta T), \\
E_i(T) &= E_i\, f_{E,i}(\Delta T),
\end{split}
\end{equation}
with $\alpha_i(T)$ the \emph{instantaneous} coefficient of thermal expansion. 

\begin{figure*}[h]
\centering
\includegraphics[width=\textwidth]{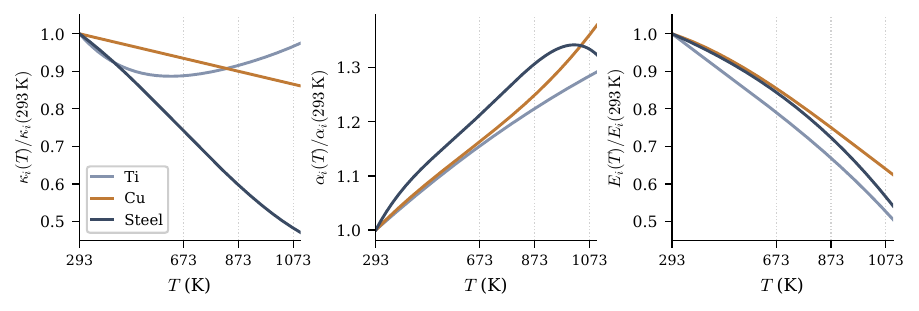}
\caption{Fitted temperature-dependent property factors for Ti, Cu, and Steel over the validity window $293\text{--}1100$ K, normalized by the room-temperature values given in the text: thermal conductivity (left), instantaneous coefficient of thermal expansion (center), and Young's modulus (right). Dotted lines mark the three design temperatures.}
\label{fig:props}
\end{figure*}

The factors in Fig.~\ref{fig:props} are constrained least-squares fits, exact at 293 K, to recommended data: TPRC/Touloukian conductivity and expansion series for Ti and Cu \cite{ho1972, touloukian1975}, plain carbon steel conductivity \cite{incropera2007}, and elastic moduli from single-crystal and polycrystal measurements for Ti \cite{fisher1964}, Cu \cite{chang1966}, and Steel \cite{dever1972}; the fits reproduce the anchors to within $0.4\%$. The Curie point of Steel ($\approx 1043$ K) lies inside the window and is smoothed through by the fits, and published Steel expansion data disagree by $10\text{--}15\%$ above 600 K, an irreducible uncertainty band inherited by $\epsth$ for Steel. Outside the validity window the factors are extrapolated as constants with vanishing temperature derivatives, which prevents the unguarded polynomials from turning negative and guarantees a bounded thermal energy.
%===================================================================
%===================================================================
The thermal eigenstrain uses the exact integral of the instantaneous coefficient, cf.\ Eq.~\eqref{eq:propfactor},
\begin{equation}\label{eq:eigenstrain}
\epsth(\bm{x}, T) = \sum_{i} \rho_i(\bm{x})\, \alpha_i \big[F_{\alpha,i}(\Delta T) - F_{\alpha,i}(\Delta T_\infty)\big],
\end{equation}
where $F_{\alpha,i}$ is the antiderivative of $f_{\alpha,i}$; for constant factors this reduces identically to $\alpha (T - \Tinf)$. The fitted coefficients in \eqref{eq:propfactor} and the expansion antiderivatives entering \eqref{eq:split} are tabulated in Appendix \ref{app:props}. The \emph{constant-property} model used in the comparisons anchors the properties at the design temperature: it evaluates $\kappa_i$ and $E_i$ at $\TD$ and sets $\epsth = \sum_i \rho_i\, \bar{\alpha}_i(\TD)\, (T - \Tinf)$, where
\begin{equation}\label{eq:secant}
\bar{\alpha}_i(\TD) = \frac{\alpha_i \big[F_{\alpha,i}(\Delta T_D) - F_{\alpha,i}(\Delta T_\infty)\big]}{\TD - \Tinf}
\end{equation}
is the secant (mean) expansion coefficient over $[\Tinf, \TD]$, so that the two property models coincide exactly for material at $T = \TD$. This is the stronger of the two constant-property conventions used in practice; the room-temperature convention adopted in much of the literature \cite{deng2017, sun2026} ($\kappa_i$, $E_i$, $\alpha_i$ evaluated at $293$ K) trivially misrepresents the thermal load itself. Within the optimizer, we treat the property--temperature couplings in the same Picard fashion as Eq.~\eqref{eq:heat}, while we consistently carry the design pathways ($\partial \mathcal{P}/\partial \boldsymbol{\rho}$ at frozen $T$) and the property-derivative pathways ($\partial \kappa/\partial T$, $\partial E/\partial T$, and the instantaneous $\alpha$ entering the thermo-mechanical coupling source) into the adjoint problems so that the sensitivities correspond to the clamped, temperature-dependent model actually solved.

\FloatBarrier
\subsection{Simultaneous Optimization Framework}\label{sec:framework}

The designs are produced with the m-PIGP framework, which augments our multi-material PIGP approach \cite{sun2026} with manufacturability constraints; we summarize it briefly below and extend it to the new physics, referring the reader to Appendix~\ref{app:mpigp} for details on field parameterization as well as constraint and loss formulations. 

As shown in Figure~\ref{fig:flowchart}, the primal fields ($\bm{u}$, $T$), the temperature adjoint $\mu$, and the design field $\boldsymbol{\rho}$ are parameterized by independent Gaussian process (GP) priors whose mean functions are parametric grid convolutional attention networks (PGCANs) \cite{shishehbor2024}; conditioning the GPs on boundary data enforces Dirichlet BCs exactly \cite{mora2025, yousefpour2025}, and a softmax head enforces the partition of unity on $\boldsymbol{\rho}$. All parameters are trained simultaneously with Adam \cite{kingma2014} on a loss that combines the (adjoint-augmented) design objective, the deep-energy-method residuals ($\Pi_u$, $\Pi_T$, and $\Pi_\mu$ of the primal and adjoint problems), and the design constraints; the full loss $\mathcal{L}$ is assembled in Eq.~\eqref{eq:loss} of Appendix~\ref{app:mpigp}. The continuous fields are evaluated on a single uniform $200\times100$ grid of bilinear quadrilaterals with full $2\!\times\!2$ Gauss quadrature that hosts the energies, design field, sensitivities, and adjoint solve to eliminate inter-grid transfer operators.

The design sensitivities that drive the update of $\boldsymbol{\rho}$ are obtained from an adjoint analysis that accounts for the finite-strain kinematics and for all property--temperature couplings. Since both aspects differ substantially from the small-strain, constant-property adjoint of our previous work, we derive them in Sec.~\ref{sec:sens}.

We enforce manufacturability by two mechanisms that are detailed in  Appendix~\ref{app:mpigp}: a Helmholtz PDE filter \cite{lazarov2011} with radius $r = 5\,\mu$m controls the minimum feature size and suppresses islands, and a pairwise interface-exclusion penalty prevents direct Ti--Steel contact, which forms brittle Ti--Fe intermetallics \cite{moshokoa2024,jafari2025}. The mass constraint limits the design to $25\%$ of the mass of a fully dense copper domain and is ramped over the first half of training together with the SIMP exponent ($p: 1 \to 3$); the interface penalty follows its own continuation schedule (Appendix~\ref{app:mpigp}). 

We run each course of optimization for $10{,}000$ epochs with five independent initializations. For each device, we set the design temperature to $\TD \in \{673, 873, 1073\}$ K and specify the physics model (baseline: linear elasticity with $\TD$-anchored constant properties; high-fidelity: Hencky kinematics with temperature-dependent properties). This procedure produces $60$ optimized designs. 

\FloatBarrier
\subsection{Adjoint Sensitivity Analysis}\label{sec:sens}
The stroke objective depends on the design only through the coupled state problem, which is nonlinear in both $\bm{u}$ (finite strain) and $T$ (temperature-dependent conductivity). We let $\bm{R}_T(\mathbf{T}; \boldsymbol{\rho}) = \bm{0}$ and $\bm{R}_u(\mathbf{u}; \mathbf{T}, \boldsymbol{\rho}) = \partial \Pi_u / \partial \mathbf{u} = \bm{0}$ for the discrete thermal and mechanical residuals on the unified grid, with $\Pi_u$ from Eq.~\eqref{eq:potential}. The coupling is one-way: heat transfer does not depend on $\bm{u}$. We augment the objective as $\mathcal{J} = \uout - \boldsymbol{\lambda}^\top \bm{R}_u - \boldsymbol{\mu}^\top \bm{R}_T$ (distinct from the training loss $\mathcal{L}$ of Appendix~\ref{app:mpigp}) and require stationarity with respect to both states. This defines the two adjoint problems below.

\textbf{Displacement adjoint.} Stationarity with respect to $\mathbf{u}$ gives
\begin{equation}\label{eq:uadj}
\mathbf{K}_T(\mathbf{u}^*)\, \boldsymbol{\lambda} = \bm{e}_{\mathrm{out}}, \qquad
\mathbf{K}_T = \frac{\partial^2 \Pi_u}{\partial \mathbf{u}\, \partial \mathbf{u}}\bigg|_{\mathbf{u}^*}.
\end{equation}

Here $\bm{e}_{\mathrm{out}}$ is a unit load at the output port in the output direction, and $\mathbf{K}_T$ is the consistent tangent at the converged state. It contains the material part, the geometric (initial-stress) part, and the port-spring contribution $K_s$. Because the thermally pre-stressed members carry compression, $\mathbf{K}_T$ is symmetric but can be indefinite; we therefore solve Eq.~\eqref{eq:uadj} with a sparse LDL$^\top$ factorization \cite{chi2026torchsla} rather than energy minimization.
 
We assemble $\mathbf{K}_T$ from batched per-element Hessians of the interpolated energy in Eq.~\eqref{eq:wang}, using vectorized automatic differentiation. This is exact for the three-term form, and the parameterization of Eq.~\eqref{eq:closedform} keeps the second derivatives bounded even at coalescent stretches. We then impose $\boldsymbol{\lambda} = \bm{0}$ on exactly the kinematically constrained degrees of freedom of the primal problem. During training, we refresh the factorization every 25 epochs after a 2000-epoch primal warm-up and treat the adjoint field as constant (detached) between refreshes.

\textbf{Temperature adjoint.} Stationarity with respect to $\mathbf{T}$ gives $(\partial \bm{R}_T/\partial \mathbf{T})^\top \boldsymbol{\mu} = -(\partial \bm{R}_u/\partial \mathbf{T})^\top \boldsymbol{\lambda}$. Since $\kappa$ depends on $T$, the thermal tangent is \emph{nonsymmetric} and its transpose yields the weak problem: find $\boldsymbol{\mu}$ with $\mu = 0$ on the hot edge such that for all admissible $w$,
\begin{equation}\label{eq:Tadj}
\begin{split}
\int_\Omega \kappa\, \nabla w \cdot \nabla \mu\, \mathrm{d}V
&+ \int_\Omega \frac{\partial \kappa}{\partial T}\Big|_{T^*} \big(\nabla T^* \cdot \nabla \mu\big)\, w\, \mathrm{d}V \\
&= \int_\Omega \Big[ \alpha_{\mathrm{inst}}\, \mathcal{T}_\lambda - \frac{\partial_T E}{E}\, \mathcal{S}_\lambda \Big]\, w\, \mathrm{d}V,
\end{split}
\end{equation}
where $\alpha_{\mathrm{inst}} = \partial \epsth/\partial T$ is the instantaneous expansion coefficient of Eq.~\eqref{eq:propfactor}. 

We emphasize that the second term on the left never appears in the primal solve. $\kappa$ is frozen at the previous temperature guess, so each Picard step is a \emph{linear} conduction problem and never differentiates $\kappa$ with respect to $T$. The adjoint, however, linearizes the converged nonlinear residual $\bm{R}_T(\mathbf{T}) = \bm{0}$ itself, which does depend on $T$ through $\kappa(T)$, so omitting this term would transpose the Picard operator instead of the true tangent and bias every sensitivity.

The right-hand side collects the two paths through which the mechanical residual senses temperature: an eigenstrain path, from thermal expansion generating internal force, and a modulus path, from $E(T)$ softening the stiffness directly. For solid material, the eigenstrain path is
\begin{equation}\label{eq:Tpath}
\mathcal{T}_\lambda = \frac{E}{1-\nu}\, \mathbf{F}^{-\top} : \nabla \boldsymbol{\lambda},
\end{equation}
which follows from the identity $\operatorname{tr} \boldsymbol{\varepsilon}_H = \ln \det
\mathbf{F}$, whose directional derivative along $\boldsymbol{\lambda}$ is exactly
$\mathbf{F}^{-\top} : \nabla \boldsymbol{\lambda}$. This is the only change finite strain
makes to the classical small-strain trace term $\operatorname{tr} \nabla \boldsymbol{\lambda}$.
Under the interpolation of Eq.~\eqref{eq:wang}, the two branches combine as
$\gamma \mathbf{F}_\gamma^{-\top} : \nabla \boldsymbol{\lambda} + (1-\gamma)
\operatorname{tr} \nabla \boldsymbol{\lambda}$; the linear branch of the three-term form
carries weight $1-\gamma$ because the eigenstrain cross term is linear in the strain.

The modulus path exploits a different structure: every term of the interpolated energy is
linear-homogeneous in the local modulus $E(\boldsymbol{\rho}, T)$. The energy's temperature
derivative is therefore its own directional derivative along the adjoint,
$\mathcal{S}_\lambda = \mathrm{d} \psi(\mathbf{H}^* + \tau \nabla \boldsymbol{\lambda}) /
\mathrm{d}\tau |_{\tau=0}$, scaled by $\partial_T E / E$.

For constant properties, $\partial_T \kappa = \partial_T E = 0$ and $\alpha_{\mathrm{inst}}
= \bar{\alpha}(\TD)$, so Eq.~\eqref{eq:Tadj} reduces to a self-adjoint conduction problem
driven by the eigenstrain path alone. Linear kinematics then recovers the classical
small-strain framework, with the source in its $\operatorname{tr} \nabla \boldsymbol{\lambda}$
form. Under Hencky kinematics, the source retains its interpolated finite-strain form even at
constant properties.

We parametrize $\boldsymbol{\mu}$ by its own GP-conditioned network and
train it on the functional $\Pi_\mu$ whose stationarity condition is Eq.~\eqref{eq:Tadj}, with the nonsymmetric term entering in the same Picard-lagged fashion as the primal properties, i.e., $\partial_T \kappa$ is evaluated at the previous temperature iterate,
so at the fixed point the assembled operator is the true tangent.

\textbf{Design sensitivity.} With the adjoints so defined, the total derivative reduces to
partial derivatives at frozen states,
\begin{equation}\label{eq:sens}
\frac{\mathrm{d} \uout}{\mathrm{d} \rho_i} =
- \frac{\partial W_\lambda}{\partial \rho_i}
- \int_\Omega \frac{\partial \kappa}{\partial \rho_i} \nabla T^* \cdot \nabla \mu\, \mathrm{d}V
+ \int_\Omega \frac{\partial q_v}{\partial \rho_i}\, \mu\, \mathrm{d}V,
\end{equation}
where $W_\lambda(\boldsymbol{\rho}) := \mathrm{d}\, \Pi_{\mathrm{int}}(\mathbf{H}^* + \tau
\nabla \boldsymbol{\lambda};\, \boldsymbol{\rho})/\mathrm{d}\tau |_{\tau=0}$ is the
adjoint-weighted internal virtual work, evaluated with frozen states $(\mathbf{u}^*,
\boldsymbol{\lambda}, T^*)$ but design-\emph{live} coefficients.

This definition fuses the three mechanical design pathways into one object: a single
automatic-differentiation pass through $W_\lambda$ carries the stiffness path $\partial
\mathbb{C}/\partial \boldsymbol{\rho}$, the eigenstrain path $\partial \epsth / \partial
\boldsymbol{\rho}$, and the interpolation-factor path $\partial \gamma / \partial
\boldsymbol{\rho}$ simultaneously.

Finally, the same automatic
differentiation pulls the gradient of Eq.~\eqref{eq:sens} back through the Helmholtz filter
and the softmax partition of unity (Appendix~\ref{app:mpigp}); since the filter map
$(\mathbf{I} - r^2 \Delta)^{-1}$ is self-adjoint, this is equivalent to filtering the
sensitivity field itself.

\FloatBarrier
\section{Results and Discussion}\label{sec:results}

Putting a modeling assumption on trial naturally raises two distinct questions: (1) how much it distorts the predicted performance of a given topology, and (2) to what extent it affects the designed topology compared to the case where a higher-fidelity model is used. We answer the first question in Sections~\ref{sec:designs}--\ref{sec:kinematics} which establish the evaluation error at fixed design and trace it to its kinematic origin. Then, in Sec.~\ref{sec:payoff}, we measure the gain from designing with better physics, weighted against its cost. Finally, in Sec.~\ref{sec:robust}, we investigate the performance robustness of the devices away from their design point.

\textbf{Evaluation Protocol.} Comparing designs by the objective values reported by their own optimizers would conflate model bias with design quality. We therefore freeze every converged design at its categorical material map (each element assigned to void, Ti, Cu, or Steel) and re-solve it with a standalone finite element evaluator on the same grid, weak forms, and BCs, under all four physics models
\begin{equation*}
\{\text{linear}, \text{Hencky}\} \times \{\text{constant (at $\TD$)}, \text{$T$-dependent properties}\},
\end{equation*}
and at all three temperatures $\TD \in \{673, 873, 1073\}$ K, regardless of the design's training conditions. The thermal problem is solved by Picard iteration (tolerance $10^{-11}$); the linear-elastic problems by a single sparse direct solve; and the Hencky problems by incremental Newton--Raphson with the consistent tangent, ramping the thermal eigenstrain in four increments with automatic cutback. Incrementation is not cosmetic: at zero displacement under the full thermal load, the compressive thermal pre-stress renders the tangent indefinite, and Newton--Raphson iteration started from the undeformed state diverges. Convergence requires the maximum residual to drop below $10^{-11}$ of its initial value in double precision. The evaluator's constitutive routines are verified against the training implementation at the source level, and against analytical solutions by free-thermal-expansion patch tests, for which the four models give four distinct exact answers (e.g., $u = (e^{\epsth}{-}1) X$ for Hencky versus $u = \epsth X$ for linear kinematics), reproduced to a relative error below $10^{-10}$.

For each design evaluated at its own $\TD$, we report the four strokes $u(\mathcal{C}, \mathcal{P})$ with constitutive law $\mathcal{C} \in \{\text{lin}, \text{Hen}\}$ and property model $\mathcal{P} \in \{\text{const}, T\text{-dep}\}$, and define, normalizing by the reference cell $\uref = u(\text{Hen}, T\text{-dep})$:
\begin{align}
\dlaw^{\mathcal{P}} &= \big[u(\text{Hen}, \mathcal{P}) - u(\text{lin}, \mathcal{P})\big]/\uref, \label{eq:dlaw}\\
\dprop^{\mathcal{C}} &= \big[u(\mathcal{C}, T\text{-dep}) - u(\mathcal{C}, \text{const})\big]/\uref, \label{eq:dprop}
\end{align}
i.e., the constitutive-law effect at fixed property model and the property effect at fixed constitutive law; their difference across levels is the interaction $\dlaw^{T\text{-dep}} - \dlaw^{\text{const}}$. The \emph{self-assessment bias} of a design family is the difference between the stroke in its native cell (the model it was optimized with) and the reference cell. Finally, evaluating every design under the reference physics at all three temperatures quantifies \emph{temperature robustness} (stroke versus operating temperature) and \emph{transferability} (which family, defined by design constitutive law and training $\TD$, performs best at each operating temperature).

\FloatBarrier
\subsection{Optimized Layouts and Material Allocation}\label{sec:designs}

\begin{figure*}[h]
\centering
\includegraphics[width=\textwidth]{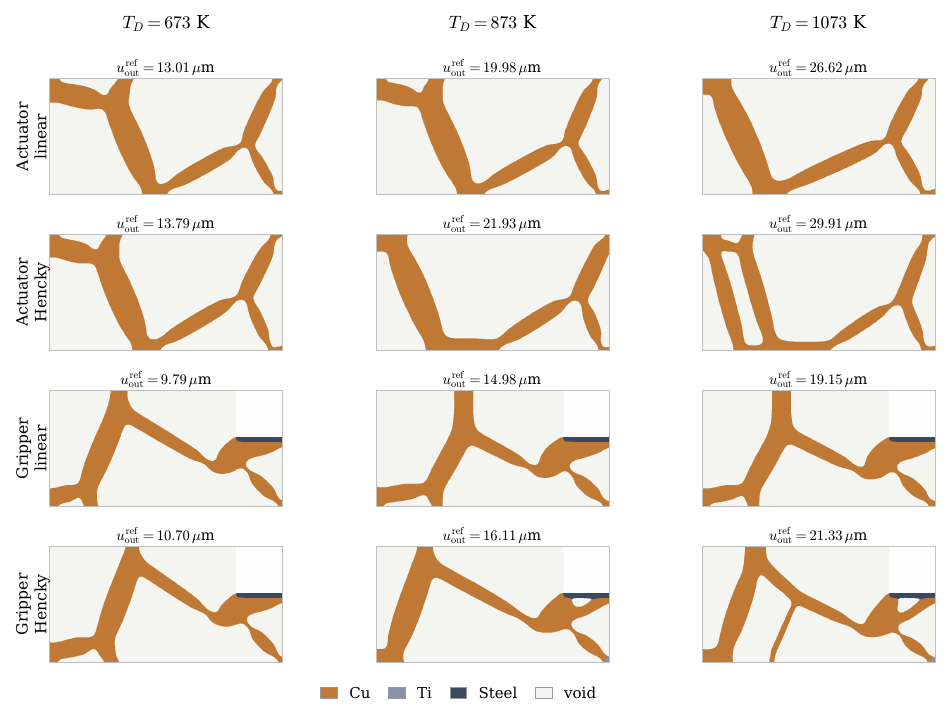}
\caption{Median-performing optimized designs (by $\uref$ among the five seeds) for the actuator (top two rows) and gripper (bottom two rows), for the baseline (linear + $\TD$-anchored constant properties) and high-fidelity (Hencky + $T$-dependent properties) design families at the three design temperatures.}
\label{fig:designs}
\end{figure*}

\begin{figure*}[h]
\centering
\includegraphics[width=\textwidth]{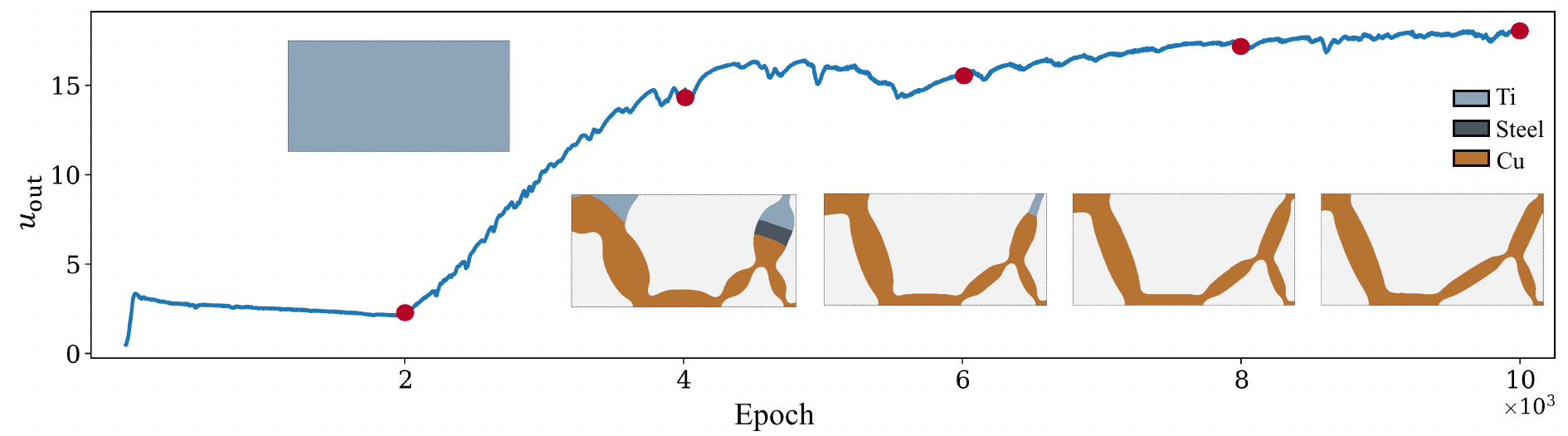}
\caption{Optimization history of a representative high-fidelity run (actuator, Hencky + $T$-dependent properties, $\TD = 873$ K): output stroke of the evolving continuous design, as estimated by the simultaneous solver, with snapshots of the material layout at the marked epochs. During the 2000-epoch primal warm-up the design field stays nearly uniform; once the displacement adjoint activates and the mass constraint, SIMP exponent, and interface penalty ramp over the first half of training, the mechanism forms rapidly, and the remaining epochs refine hinges and limb inclinations while titanium and steel are eliminated in favor of copper.}
\label{fig:evolution}
\end{figure*}

The high-fidelity and baseline design families primarily differ in their topologies rather than their overall material composition and mass. Hence, the performance discrepancies that we report throughout this section stem almost entirely from how the physics model drives the spatial material distribution. Fig.~\ref{fig:designs} shows the median-performing design of each family, Fig.~\ref{fig:evolution} instantiates how such a design emerges over a single optimization run. We summarize the reference-physics stroke and the realized mass of all sixty manufactured designs in Table~\ref{tab:massstats}.

Our main observations are three-fold.
%====================================================================
%====================================================================

%====================================================================
%====================================================================
First, the material allocation is robust to the physics upgrade. With direct Ti--Steel contact excluded, copper offers the best combination of thermal conductivity and expansion among the permitted phases so every design is almost entirely copper. In gripper design, we prescribe the steel phase at the jaw to make that design region strictly solid. Second, mass cannot explain any of the performance differences reported below. The realized masses settle within $2\%$ of the budget (marginally above it, due to the finite weight of the quadratic mass penalty and the thresholding of the continuous design field into its categorical material map), the two families use practically identical mass at every design temperature, and the per--seed mass scatter is negligible even in the families with large stroke scatter. In particular, the collapsed seeds visible in the min columns of Table~\ref{tab:massstats} are inferior \emph{mechanisms}, not lighter designs (Sec.~\ref{sec:payoff}). Third, although the material budget and the actuation archetype (i.e., angled expanding limbs that lever the output port) are common to both families, the design-time physics systematically reshapes the mechanism itself: the high-fidelity layouts differ in limb inclination and hinge placement, and at $\TD = 1073$ K the Hencky-designed actuator develops distinct closed structural loops near the hot edge that its linear counterpart never finds. Judged by the reference physics, the median high-fidelity design outperforms the median baseline design at every temperature for both devices (Fig.~\ref{fig:designs}, Table~\ref{tab:massstats}).

\FloatBarrier
%====================================================================
\begin{table}[h]
\caption{Statistics of the reference-physics stroke ($\mu$m) and of the realized mass ($\mu$g) of the manufactured (categorical) designs at their own design temperature, over the five seeds per family. The mass budget is $4.200\,\mu$g for the actuator and $3.864\,\mu$g for the gripper ($25\%$ of a fully dense copper domain).}
\label{tab:massstats}
\centering%\small
\setlength{\tabcolsep}{3.6pt}
\resizebox{\textwidth}{!}{%
\begin{tabular}{lllcccccccccc}
\toprule
 & & & \multicolumn{5}{c}{Actuator} & \multicolumn{5}{c}{Gripper} \\
\cmidrule(lr){4-8}\cmidrule(lr){9-13}
$T_D  (\text{K})$ & Constitutive Law & Metric & median & mean & min & max & std & median & mean & min & max & std \\
\midrule
\multirow{4}{*}{673}
 & \multirow{2}{*}{Linear} & $\uout^{\mathrm{ref}}$ & 13.010 & 12.222 & 8.676 & 13.395 & 1.990 & 9.795 & 9.835 & 9.736 & 9.950 & 0.096 \\
 &                         & mass & 4.227 & 4.224 & 4.204 & 4.231 & 0.011 & 3.896 & 3.896 & 3.892 & 3.899 & 0.003 \\
\addlinespace[1.5pt]
 & \multirow{2}{*}{Hencky} & $\uout^{\mathrm{ref}}$ & 13.786 & 12.820 & 8.823 & 13.939 & 2.235 & 10.701 & 10.705 & 10.494 & 10.883 & 0.146 \\
 &                         & mass & 4.230 & 4.227 & 4.213 & 4.233 & 0.008 & 3.898 & 3.897 & 3.887 & 3.909 & 0.008 \\
\midrule
\multirow{4}{*}{873}
 & \multirow{2}{*}{Linear} & $\uout^{\mathrm{ref}}$ & 19.984 & 20.083 & 19.872 & 20.510 & 0.256 & 14.980 & 13.933 & 9.622 & 15.202 & 2.413 \\
 &                         & mass & 4.238 & 4.237 & 4.232 & 4.242 & 0.004 & 3.898 & 3.897 & 3.883 & 3.908 & 0.009 \\
\addlinespace[1.5pt]
 & \multirow{2}{*}{Hencky} & $\uout^{\mathrm{ref}}$ & 21.935 & 21.930 & 21.703 & 22.273 & 0.219 & 16.107 & 15.210 & 10.707 & 17.118 & 2.569 \\
 &                         & mass & 4.236 & 4.237 & 4.229 & 4.243 & 0.006 & 3.904 & 3.905 & 3.899 & 3.911 & 0.005 \\
\midrule
\multirow{4}{*}{1073}
 & \multirow{2}{*}{Linear} & $\uout^{\mathrm{ref}}$ & 26.620 & 26.519 & 26.135 & 26.662 & 0.218 & 19.154 & 17.235 & 12.861 & 19.859 & 3.153 \\
 &                         & mass & 4.245 & 4.247 & 4.236 & 4.268 & 0.012 & 3.903 & 3.901 & 3.886 & 3.911 & 0.009 \\
\addlinespace[1.5pt]
 & \multirow{2}{*}{Hencky} & $\uout^{\mathrm{ref}}$ & 29.906 & 29.878 & 29.575 & 30.192 & 0.240 & 21.331 & 19.705 & 15.591 & 21.892 & 2.779 \\
 &                         & mass & 4.246 & 4.252 & 4.244 & 4.270 & 0.011 & 3.906 & 3.910 & 3.900 & 3.926 & 0.010 \\
\bottomrule
\end{tabular}%
}
\end{table}
%==============================================
\subsection{Evaluation-Model Error: Constitutive Law Versus Property Model}\label{sec:evalmodel}

\begin{figure*}[h]
\centering
\includegraphics[width=\textwidth]{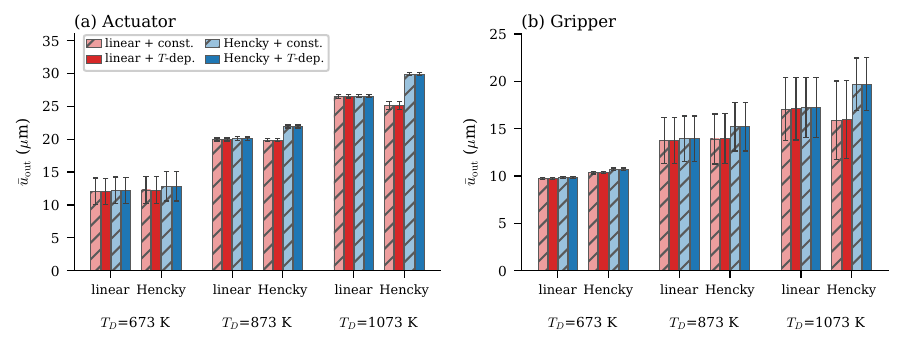}
\caption{Cross-evaluation of every design at its own design temperature: mean output stroke ($\pm$ one standard deviation over the five seeds) under the four evaluation models, grouped by design family and $\TD$, for (a) the actuator and (b) the gripper. Several families contain one or two seeds that collapsed to inferior local minima ($21\text{--}31\%$ below the family mean), which lowers their means and inflates their error bars; these collapses are discussed in Sec.~\ref{sec:payoff}.}
\label{fig:crossbars}
\end{figure*}

\begin{figure}[h]
\centering
\includegraphics[width=0.7\textwidth]{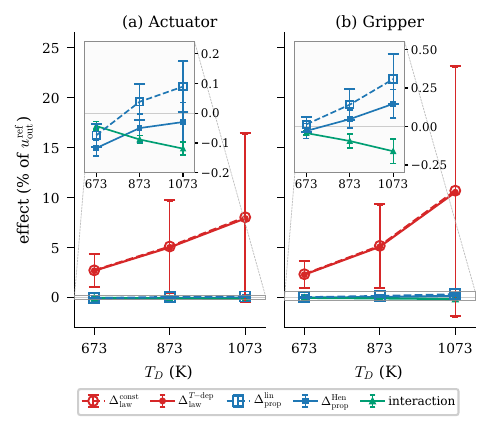}
\caption{Factorial effects of the evaluation physics at the designs' own $\TD$, in percent of the reference stroke (mean $\pm$ one standard deviation over the ten designs at each $\TD$) for (a) the actuator and (b) the gripper: with the constant properties anchored at $\TD$, the property effects and the interaction are one to two orders of magnitude smaller than the constitutive-law effect and collapse onto the zero line; the insets magnify them (same percent units, panel-specific scale). The interaction is thereby resolved from both property effects: it stays below $0.2\%$ in the mean, but is systematically negative and grows in magnitude with $\TD$ for both devices.}
\label{fig:effects}
\end{figure}

A systematic cross-evaluation of the optimal designs demonstrates that the impact of the two physics modeling assumptions is markedly disproportionate. The material property model, whether the properties presumed are uniform or spatially varying, affects the performance negligibly compared to the choice of constitutive law. We show the mean stroke of each design family in the four combinations of assumptions at its own design temperature in Fig.~\ref{fig:crossbars}, and Fig.~\ref{fig:effects} condenses them into the factorial effects of Eqs.~\eqref{eq:dlaw}--\eqref{eq:dprop}, which separate the two assumptions by an order of magnitude.
%===========================================================
%===========================================================
The constitutive-law effect is the dominant modeling error, and it is both strongly temperature- and design-dependent. Switching the evaluator from linear to Hencky kinematics raises the predicted stroke from $2\text{--}3\%$ at 673 K to $8\text{--}11\%$ at $1073$ K on average (Fig.~\ref{fig:effects}). The average metric, however, masks the nature of this error. The large spread at high temperature is not noise but systematic design-to-design variability: at $1073$ K, the high-fidelity families are misjudged by an average of $16\%$ (actuator) and $19\%$ (gripper) of the family stroke, compared to about $1\%$ for the baseline families. What the error tracks is the compound effect of rotation content and performance. Designs optimized with the Hencky model exploit finite-rotation kinematics that linear analysis cannot represent, and how badly a layout is misjudged correlates with how much rotation it carries. Because rotation and maximum stroke do not perfectly align within a family, extreme values like the $34\%$ case in Fig.~\ref{fig:hmax} represent outliers rather than the norm. Section~\ref{sec:kinematics} traces the kinematic roots of these inaccuracies.

Once we anchor the constant properties at the design temperature, the property model barely matters. In Fig. \ref{fig:crossbars}, the constant-property bar and the $T$-dependent bar of the same constitutive law are indistinguishable in every family and at every design temperature. The insets of Fig. \ref{fig:effects} magnify the difference: the mean property effect never exceeds $0.31\%$ of the reference stroke, and even the error-bar extremes stay below half a percent. The physical reason is that these conduction-driven devices feature a hot edge held at $T_D$ and a mild distributed heat sink so the temperature falls no more than $\sim 60$ K below $T_D$. Since the secant expansion coefficient is exact at $T_D$ by construction, the properties frozen at the driving temperature track the solution of spatially varying $T$-dependent properties with sufficient accuracy. Therefore, selecting an appropriate anchoring point for our designs rather than relying on the common room-temperature convention \cite{sun2026,deng2017} justifies the assumption of uniform material properties across the domain. 

The two effects are also practically independent, though not exactly orthogonal. The interaction magnitude stays below $0.25\%$ of the reference stroke (see Fig. \ref{fig:effects} insets), so we can assess and adopt the kinematic upgrade separately from the property model. A $\TD$-anchored linear model consequently inherits the full constitutive-law error and the property model adds negligible error.

The interaction is nevertheless systematically negative and it encodes a real disparity
between the two constitutive laws. By Eqs.~\eqref{eq:dlaw}--\eqref{eq:dprop}, the
interaction equals $\Delta^{\mathrm{Hen}}_{\mathrm{prop}} -
\Delta^{\mathrm{lin}}_{\mathrm{prop}}$: a negative value means the linear evaluator is the
more property-sensitive of the two. This gap widens with temperature, falling from
$-0.04\%$ at 673 K to $-0.12\%$ (actuator) and $-0.16\%$ (gripper) at 1073 K; Sec.~\ref{sec:kinematics}
explains its sign. The linear model stores thermal expansion as elastic strain, whereas the Hencky model releases it as rotation. Because elastic strain energy is governed by the moduli, the linear prediction becomes more exposed to the property model. At a tenth of a percent, this coupling has no practical consequence but it marks one place in the factorial where the two assumptions measurably interact.
%===========================================================
%===========================================================
\FloatBarrier
\subsection{Kinematic Origin of the Constitutive-Law Error}\label{sec:kinematics}
These thermal devices deform as linkages: nearly rigid limbs swinging about compliant hinges, through angles large enough that the small-strain measure charges the rotation itself as a spurious compressive strain comparable to the thermal eigenstrain that drives the device. This kinematic error explains the trends observed in Sec.~\ref{sec:evalmodel}: why $\dlaw$ grows with temperature, and why it falls hardest on the layouts that rotate the most.

\begin{figure*}[h]
\centering
\includegraphics[width=\textwidth]{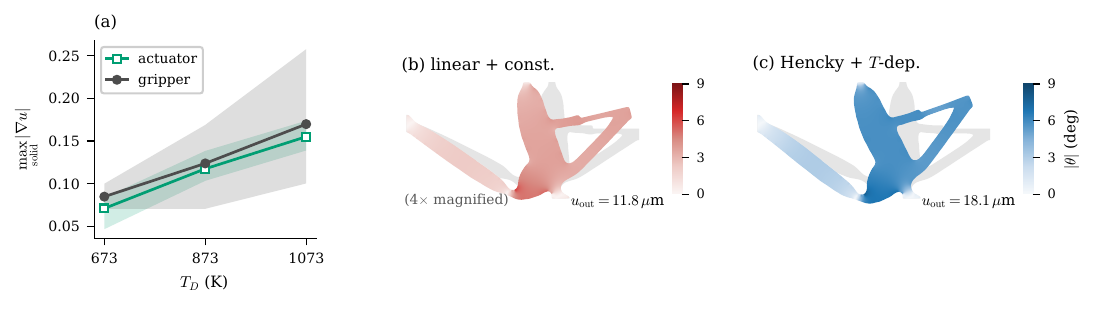}
\caption{Kinematic origin of the constitutive-law effect: (a) maximum displacement-gradient magnitude in solid material under the reference physics versus design temperature, for the designs evaluated at their own $\TD$; (b,c) deformed configuration ($4\times$ magnified, undeformed outline in gray) of the Hencky-designed gripper with the largest $\dlaw$ ($\TD = 1073$ K, seed 1), predicted by the baseline evaluation model (b) and the reference model (c), color coded by the local rigid-rotation angle $|\theta|$ from the polar decomposition $\mathbf{F} = \mathbf{R}\mathbf{U}$. The linear, constant-property evaluation underpredicts this design's stroke by $34\%$. Each limb rotates almost rigidly while $\theta$ changes abruptly across the compliant hinges; under linear kinematics this rotation produces a spurious compressive strain that is comparable to the thermal eigenstrain driving the device.}
\label{fig:deformed}
\label{fig:hmax}
\end{figure*}

Under the reference physics, the maximum displacement gradient in the solid grows almost linearly with the design temperature (Fig.~\ref{fig:hmax}(a)), while the elastic logarithmic strain stays at the percent level away from the hinges. This gap tells us the growth is coming from rotation-dominated deformation. Writing the deformation gradient as $\mathbf{F} = \mathbf{RU}$, a rotation $\mathbf{R}$ times a stretch $\mathbf{U}$, the Hencky strain depends on the stretch alone, $\varepsilon_H = \ln\mathbf{U}$ with no small-angle assumption involved. Holding $\varepsilon_H$ near 1\% therefore pins $\mathbf{U}$ within about 1\% of the identity, which means
$\mathbf{F}$ itself is within about 1\% of a pure rotation.

We make this concrete by presenting the most severely misjudged design in Figures~\ref{fig:hmax}(b,c) and color code the deformed configurations by the local rigid-rotation angle $\theta$ obtained from the polar decomposition $\mathbf{F} = \mathbf{R}\mathbf{U}$. Each limb rotates almost rigidly by up to ${\sim}8^\circ$ and $\theta$ changes abruptly across the compliant hinges that connect the limbs to the frame and accommodate rotation. Here, we split the displacement gradient as $\mathbf{H} = \mathbf{F}-\mathbf{I} = \mathbf{R}(\mathbf{U}-\mathbf{I}) + (\mathbf{R}-\mathbf{I})$ to isolate the same $\sim$1\% strain piece from a rotation piece of size $\|\mathbf{R}-\mathbf{I}\|_2 = 2\sin(\theta/2)$. This comes to roughly 14\%, an order of magnitude larger than the strain piece.

This is where linear kinematics fails. The small-strain measure assigns strain energy to the symmetric part of the displacement gradient. This linearization is exact only to first order in the rotation: a pure rotation by $\theta$ carries the spurious compressive normal strain $\cos\theta - 1 \approx -\theta^2/2$, which at the observed hinge rotations is comparable to the thermal eigenstrain ($\epsth \approx 1.5\%$) that drives the entire device, whereas the logarithmic measure correctly assigns it zero strain. The linear model artificially stiffens the hinges and their connecting limbs, which restricts hinge rotation as seen in Fig.~\ref{fig:hmax}(b). Part of the thermal expansion is consequently absorbed as spurious elastic strain instead of being released into output stroke. Since the Hencky-designed layouts deliberately develop such rotating members (Sec.~\ref{sec:designs}), the linear evaluator penalizes them disproportionately.
%===========================================================
%======================================================
\FloatBarrier
\subsection{Design-Time Payoff and Cost}\label{sec:payoff}

\begin{figure}[h]
\centering
\includegraphics[width=0.6\textwidth]{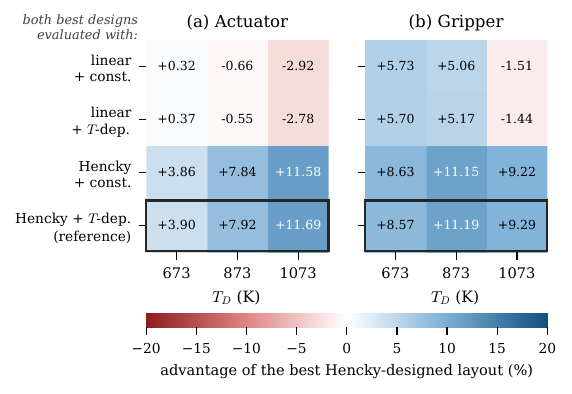}
\caption{Relative advantage of the best design of the high-fidelity family over the best design of the baseline family. Each row applies one evaluation model to \emph{both} best designs. Positive (blue) favors Hencky-designed layouts. The top rows show the self-favoring bias of the linear judges; the boxed bottom row is the reference physics, the most faithful model in the study.}
\label{fig:judges}
\end{figure}

We quantified our evaluation error in Sections~\ref{sec:evalmodel} and \ref{sec:kinematics}; now we turn to whether the upgraded physics actually produces better designs. Since a practitioner ultimately carries forward the single best design a TO routine produces, we compare the best seed of each family, with both selection and evaluation performed by one common judge. Fig.~\ref{fig:judges} reports the outcome under all four judges. The verdict of the reference judge (boxed row) is clear: the Hencky-designed layout wins all six device--temperature combinations, by $4\text{--}12\%$ of stroke. This margin is pure design-time payoff: we judge both candidates by the same physics, so the evaluation bias discussed in Sec.~\ref{sec:evalmodel} is effectively factored out.

In the rows above the reference, we illustrate what happens when the judge itself is biased. At 673 and 873 K, the gripper's advantage is large enough to survive even a linear referee. At 1073 K it is not: both linear judges rank the baseline design first on both devices, so a practitioner who screens candidates with a linear tool discards the better design precisely where the payoff is largest. This highlights why we should treat the self-assessment numbers with caution. Since the baseline families misjudge their own optimal designs by only about one percent (see Fig.~\ref{fig:crossbars}), validating a linear model against its own results creates a false sense of reliability. It appears trustworthy precisely because it avoids the rotation-rich layouts that would expose its fundamental limitations (Sec.~\ref{sec:kinematics}).
 
The payoff is also cheap to obtain. On identical single-GPU hardware (NVIDIA RTX 4090), a complete $10{,}000$-epoch optimization takes $554 \pm 7$~s for the actuator and $647 \pm 17$~s for the gripper under the baseline physics, versus $788 \pm 11$~s and $872 \pm 14$~s under the full physics (mean $\pm$ one standard deviation over the five seeds). These cost factors of $1.42$ and $1.35$ are essentially independent of $\TD$. This modest overhead arises because the simultaneous formulation never re-solves the state problem to convergence within a design step: the finite-strain model adds the three-term energies of Eq.~\eqref{eq:wang} and periodic tangent factorizations for the direct adjoint, bypassing a nested Newton loop around every analysis. Thus, critical design-time fidelity adds about forty percent computational cost. 

A final observation concerns the collapsed seeds visible in Fig.~\ref{fig:crossbars} and in the min columns of Table~\ref{tab:massstats}. These weak designs are not an artifact of finite-strain physics. They are seed-correlated across families: the same initializations collapse for the baseline and the high-fidelity frameworks alike on the same devices and design temperatures. This indicates that the rough optimization landscape is intrinsic to the simultaneous thermo-mechanical TO problem itself and worsens at higher temperatures. Thus, multi-start optimization is essential for both families. 

%======================================================
%======================================================
%===========================================================

\FloatBarrier
\subsection{Temperature Robustness and Transferability}\label{sec:robust}

\begin{figure*}[h]
\centering
\includegraphics[width=\textwidth]{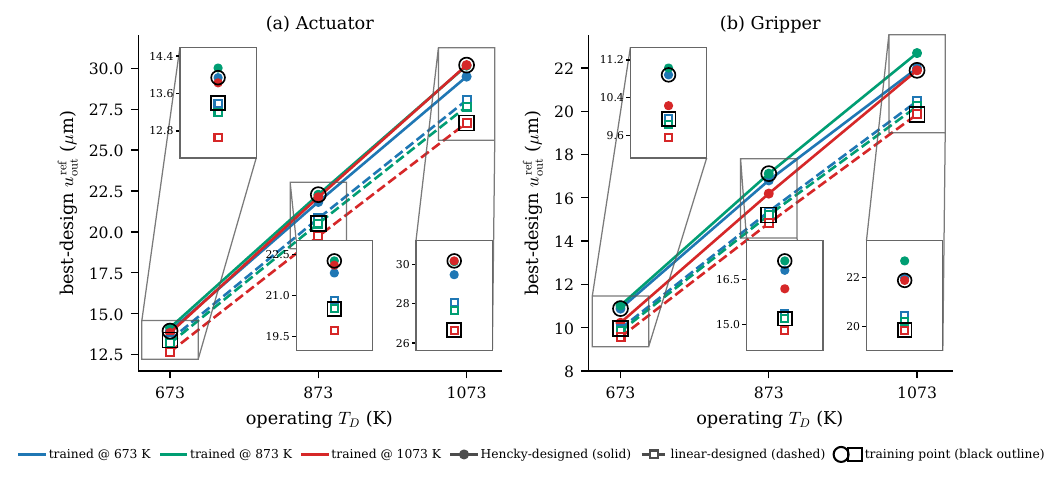}
\caption{Temperature transferability of the best designs: the best design of every family (its best seed, selected by the reference-physics stroke at its own training temperature, cf.\ Fig.~\ref{fig:judges}) evaluated under the reference physics (Hencky, $T$-dependent properties) at all three operating temperatures. The insets magnify the ordering of the best designs at each operating temperature. Hencky designs optimized at moderate temperatures are highly robust. They dominate the performance even at off-design operating temperatures.}
\label{fig:robust}
\end{figure*}

A device rarely operates at exactly its intended design temperature. To assess the robustness of the optimal designs obtained at a specific $T_D$ against off-design thermal conditions, we re-evaluate the (best) designs on driving temperatures not seen during training using reference physics (i.e., Hencky constitutive law) and report the output stroke in Fig.~\ref{fig:robust}.

Every best design transfers gracefully to off-design temperatures: the stroke is nearly affine in the operating temperature, with mid-range curvatures one to two orders of magnitude smaller than the end-to-end stroke spans. The rankings among the best designs are consequently almost independent of the operating temperature, so a design need not be operated where it was trained.

Moreover, the ranking is topped by Hencky-designed layouts in all six device--temperature combinations and, notably, not by those trained at the highest temperature. The Hencky design trained at $873$ K is the best, or within a fraction of a percent of the best in every combination; for the gripper it outperforms even the $1073$ K-trained Hencky design at $1073$ K itself. The best-design view shows that this is a statement about attainable optima rather than about averages: the deficit of the gripper family trained at $1073$ K is not caused by its weaker seeds, since even the best of its five runs falls short of the best $873$ K-trained design at its own operating temperature. The rougher high-temperature training landscape thus degrades the optima themselves. The practical guideline that follows is to optimize with the full physics at a \emph{moderate} design temperature and rely on the near-affine temperature response, rather than re-optimizing at every operating point with a less reliable high-temperature run. 
%===============================================
%===============================================
\FloatBarrier
\section{Conclusion}\label{sec:conclusion}
This paper examines two standard simplifying assumptions in thermo-mechanical topology optimization: small-strain kinematics and constant material properties. We develop a simultaneous physics-informed framework to include temperature-dependent, thus spatially varying material properties for a \{Ti, Cu, Steel\} system and a plane-stress quadratic-Hencky constitutive model, where thermal eigenstrain acts as an exact additive offset in logarithmic strain space. We account for manufacturability constraints by discouraging the Ti--Steel interface and applying a Helmholtz-type filter to the density field. We then re-evaluated all sixty converged designs using verified finite element solvers across a full factorial of constitutive laws and property models (720 solves in total). To ensure a fair comparison, the constant-property baseline was anchored at the design temperature $T_D$. 

The results reveal a sharp asymmetry between the two assumptions. These devices function as linkages with stiff limbs pivoting about compliant hinges, and linear kinematics charges the rotation itself as a spurious compressive strain comparable to the thermal eigenstrain that drives the device: its error grows from $2\text{--}3\%$ of stroke at $673$ K to $8\text{--}11\%$ at $1073$ K. For the layouts that rely most heavily on rotation, this error consumes up to a third of the stroke and inverts the ranking of the best designs at high temperatures. In contrast, the constant-property model that is anchored at the design temperature alters the results by less than one percent. However, we note that the constant-property assumption would induce higher errors if anchored at a less physically relevant temperature, or if the devices were subjected to temperature gradients significantly larger than those present in this study.

We therefore offer correspondingly simple practical guidance. We anchor constant properties at the design temperature and adopt finite-strain kinematics whenever the mechanism relies on rotation: in the simultaneous framework the upgrade replaces one energy density in the training loss rather than a solver; it costs a factor of about $1.4$ in design time, returns $4\text{--}12\%$ more best-design stroke and produces the most temperature-robust devices with moderate-temperature designs transferring best along the nearly affine stroke--temperature response. Lastly, since a linear model misjudges its own designs by only about one percent while misjudging the best designs by far more, design tools should be audited by independent higher-fidelity re-evaluation rather than by self-assessment.

The extensions we consider most pressing address the physics this study deliberately held fixed. First, material nonlinearity: at high homologous temperatures (copper at $1073$ K operates near $0.8$ of its melting point) metals yield, creep and relax, and the quadratic-Hencky energy is precisely the elastic core of multiplicative finite-strain inelasticity so the framework extends naturally to plasticity- and creep-aware design where we should consider stress and stability constraints. Second, transient response: replacing steady conduction with transient thermal loading and Joule-heating coupling would let the optimizer shape actuation bandwidth and cyclic behavior rather than a single hot state. This is particularly important for capturing the regime in which creep and thermal ratcheting accumulate over repeated strokes. Finally, optimizing over a range of operating temperatures, rather than at a single design point, would maximize expected or worst-case performance across an operating window
and yield devices that tolerate uncertain and drifting thermal environments by construction. The transferability results of this study, where moderately hot training temperatures already produce
the strongest all-around designs, are an encouraging starting point. These steps would elevate the framework from a controlled model trial to robust design of  dynamically actuated devices in service.

\FloatBarrier
\section*{Acknowledgments}
We appreciate the support from the Office of Naval Research (award number N000142312485) and National Science Foundation (award number 2238038).
\FloatBarrier
\appendix
\section{m-PIGP Training Loss and Manufacturability Constraints}\label{app:mpigp}
This appendix collects the components of the m-PIGP framework that are shared by both design families and referenced from Secs.~\ref{sec:framework} and \ref{sec:sens}: the GP parameterization of the fields, the PGCAN mean functions, the total training loss, the Helmholtz filter, the pairwise interface-exclusion penalty, and the mass constraint with its continuation schedules. Together with the physics and adjoint formulations of Sec.~\ref{sec:method}, these definitions make the paper self-contained.

\FloatBarrier
\subsection{GP Field Parameterization with Exact Boundary Conditions}\label{app:param}
The primal fields ($\bm{u}$, $T$), the temperature adjoint $\mu$, and the design field $\boldsymbol{\rho}$ are each assigned an independent GP prior. For the multi-output fields (e.g., the displacement vector and the $(n_m{+}1)$-phase design field) we follow \cite{sun2026} and pair independent kernels across outputs with a single shared multi-output mean function, which captures inter-output dependencies through the mean while keeping the conditioning of each output numerically robust and inexpensive. The displacement adjoint $\boldsymbol{\lambda}$ is deliberately \emph{not} parameterized this way: it is a solved field, obtained from the sparse direct factorization of Eq.~\eqref{eq:uadj}.

Taking the displacement field as an example, the field evaluated at an arbitrary query point $\bm{x}^* \in \Omega$ is the GP posterior mean conditioned on the prescribed boundary data:
\begin{subequations}\label{eq:gp}
\begin{align}
u_i(\bm{x}^*; \boldsymbol{\theta}_u, \phi_u)
&= m_{u_i}(\bm{x}^*; \boldsymbol{\theta}_u) + \bm{w}_{u_i}^\top \bm{r}_{u_i}, \label{eq:gp_a}\\
\bm{w}_{u_i}
&= \bm{c}_u^{-1}(\bm{x}_{u_i}, \bm{x}_{u_i}; \phi_u)\, \bm{c}_u(\bm{x}_{u_i}, \bm{x}^*; \phi_u), \label{eq:gp_b}\\
\bm{r}_{u_i}
&= \tilde{u}_i - m_{u_i}(\bm{x}_{u_i}; \boldsymbol{\theta}_u), \qquad i = 1, 2, \label{eq:gp_c}
\end{align}
\end{subequations}
where $m_{u_i}(\cdot; \boldsymbol{\theta}_u)$ is the shared mean function (Sec.~\ref{app:pgcan}), $\bm{c}_u(\cdot,\cdot; \phi_u)$ is the kernel of output $i$, and $\tilde{u}_i(\bm{x}_{u_i})$ collects the displacements prescribed at the kinematically constrained boundary points $\bm{x}_{u_i}$. Setting $\bm{x}^* = \bm{x}_{u_i}$ in Eq.~\eqref{eq:gp} recovers $\tilde{u}_i$ regardless of the mean function or kernel, so the reconstructed field satisfies its Dirichlet BCs \emph{exactly} by construction and no boundary-condition penalty appears in the loss of Eq.~\eqref{eq:loss} \cite{mora2025}. The temperature field ($T = \TD$ on the hot edge), the temperature adjoint ($\mu = 0$ on the hot edge, cf.\ Eq.~\eqref{eq:Tadj}), and the design field are parameterized analogously; for the design field, the same conditioning mechanism prescribes fixed phases in non-design regions, which is how the solid-steel gripper jaw (Sec.~\ref{sec:designs}) is enforced identically for both design families.

All GPs use the Gaussian kernel
\begin{equation}\label{eq:kernel}
c(\bm{x}, \bm{x}'; \phi) = \exp\!\big[-(\bm{x}-\bm{x}')^\top \mathrm{diag}(\phi)\,(\bm{x}-\bm{x}')\big],
\end{equation}
with a jitter $\delta = 10^{-5}$ added to the diagonal of each covariance matrix for numerical stability. The kernel hyperparameters $\phi$ are fixed to $10^{1/4}$ rather than trained \cite{mora2025, sun2026}, so the covariance matrix of each output over its conditioning points is Cholesky-factorized once at the start of training and cached; conditioning thereafter costs one cached triangular solve and one kernel--vector product per epoch, and no covariance matrix is re-factorized inside the training loop.

\FloatBarrier
\subsection{PGCAN Mean Functions}\label{app:pgcan}
The mean function of every GP is a parametric grid convolutional attention network (PGCAN) \cite{shishehbor2024}, an encoder--decoder architecture that mitigates the spectral bias of standard multi-layer perceptrons and resolves the sharp gradients and localized features (hinges, phase boundaries) on which TO depends. The encoder embeds the design domain in a trainable feature tensor $\mathbf{F}_0 \in \mathbb{R}^{N_{\mathrm{rep}} \times N_f \times N_x^e \times N_y^e}$, where $N_{\mathrm{rep}}$ grid repetitions with small diagonal offsets each carry $N_f$ features on an $N_x^e \times N_y^e$ vertex grid, and a $2\times 2$ convolution produces a feature map $\mathbf{F}_c$ of the same size. A query point $\bm{x}$ is mapped to the local coordinates $\bar{\bm{x}} \in [0,1]^2$ of its enclosing cell and cosine-transformed, $\tilde{\bm{x}} = \tfrac{1}{2}(1 - \cos \pi \bar{\bm{x}})$, and its feature vector is bilinearly interpolated from the enclosing vertices of each repetition $m$,
\begin{equation}\label{eq:bilinear}
\begin{split}
\mathbf{f}_m(\bm{x}) ={}& (1-\tilde{x})(1-\tilde{y})\, \mathbf{f}_m^{(0,0)} + (1-\tilde{x})\,\tilde{y}\, \mathbf{f}_m^{(0,1)} \\
&+ \tilde{x}(1-\tilde{y})\, \mathbf{f}_m^{(1,0)} + \tilde{x}\tilde{y}\, \mathbf{f}_m^{(1,1)},
\end{split}
\end{equation}
where $\mathbf{f}_m^{(i,j)} \in \mathbb{R}^{N_f}$ are the features of $\mathbf{F}_c$ at the vertices of the enclosing cell, and the repetitions are summed, $\mathbf{f}(\bm{x}) = \sum_{m=1}^{N_{\mathrm{rep}}} \mathbf{f}_m(\bm{x})$. Because $\mathbf{f}(\bm{x})$ depends only on the trainable features in the vicinity of $\bm{x}$, training is localized and high-frequency solution content is captured without deepening the network \cite{shishehbor2024}. The decoder splits $\mathbf{f}(\bm{x})$ into two halves that sequentially modulate the hidden states of a shallow three-layer network through PGCAN's attention mechanism, which improves gradient propagation during training.

Each field has its own PGCAN instance with separate parameters: the displacement network carries two output heads (the shared mean of the two displacement GPs), the temperature and temperature-adjoint networks carry scalar heads, and the design network carries $n_m + 1$ heads whose outputs pass through a softmax so that the mean function itself yields nonnegative fractions satisfying the partition of unity of Sec.~\ref{sec:method}. All runs use $N_f = 64$ features, an encoder resolution of $(N_x^e, N_y^e) = (48, 24)$, $N_{\mathrm{rep}} = 12$ grid repetitions, and decoders with three hidden layers of $N_f/2 = 32$ neurons.

\FloatBarrier
\subsection{Total Training Loss}\label{app:loss}
All parameters $\boldsymbol{\theta} = (\boldsymbol{\theta}_u, \boldsymbol{\theta}_T, \boldsymbol{\theta}_\mu, \boldsymbol{\theta}_\rho)$ of the PGCAN-parameterized GP priors in Fig.~\ref{fig:flowchart} are updated simultaneously by Adam on
\begin{equation}\label{eq:loss}
\mathcal{L} = \omega_c\, \mathcal{J}^{(a)}(\boldsymbol{\rho})
+ \omega_u \Pi_u + \omega_T \big(\Pi_T + \Pi_\mu\big)
+ \omega_m C_M^2 + \omega_{\mathrm{if}}\, \Pi_{\mathrm{if}}(\boldsymbol{\rho}).
\end{equation}
Here $\Pi_u$ is the mechanical potential of Eq.~\eqref{eq:potential}, $\Pi_T$ the conduction functional of Eq.~\eqref{eq:PiT}, and $\Pi_\mu$ the temperature-adjoint functional whose stationarity condition is Eq.~\eqref{eq:Tadj}; minimizing them over $(\boldsymbol{\theta}_u, \boldsymbol{\theta}_T, \boldsymbol{\theta}_\mu)$ enforces the primal and adjoint problems in deep-energy form. The adjoint-augmented objective $\mathcal{J}^{(a)}$ is a functional of the design alone: it is evaluated with frozen states $(\mathbf{u}^*, T^*)$ and frozen adjoints $(\boldsymbol{\lambda}, \boldsymbol{\mu})$ but design-live coefficients, and is assembled from the three terms of Eq.~\eqref{eq:sens} so that $\partial \mathcal{J}^{(a)}/\partial \boldsymbol{\rho} = -\,\mathrm{d}\uout/\mathrm{d}\boldsymbol{\rho}$; its minimization therefore maximizes the stroke through exactly the sensitivities derived in Sec.~\ref{sec:sens}. The displacement adjoint $\boldsymbol{\lambda}$ is obtained from the direct solve of Eq.~\eqref{eq:uadj} and enters Eq.~\eqref{eq:loss} as a constant between refreshes. The weights are $\omega_c = \omega_u = \omega_T = 10^3$ and $\omega_m = 10^5$; $\omega_{\mathrm{if}}$ follows the continuation schedule of Sec.~\ref{app:interface}. Gradients with respect to $\boldsymbol{\theta}_\rho$ are back-propagated through the softmax partition of unity and through the Helmholtz filter described next.

\FloatBarrier
\subsection{Helmholtz PDE Filter}\label{app:filter}
Let $\tilde{\boldsymbol{\rho}}(\bm{x})$ denote the raw design field produced by the design network and GP conditioning. The physical field $\boldsymbol{\rho}$ that enters all state, adjoint, and penalty evaluations is obtained per phase from the Helmholtz PDE \cite{lazarov2011}
\begin{equation}\label{eq:filter}
\big(\mathbf{I} - r^2 \Delta\big)\, \boldsymbol{\rho}(\bm{x}) = \tilde{\boldsymbol{\rho}}(\bm{x}), \qquad \nabla \boldsymbol{\rho} \cdot \bm{n} = \bm{0} \ \text{on } \partial\Omega,
\end{equation}
with radius $r = 5\,\mu$m. In the Fourier domain a mode of wavenumber $\bm{k}$ is attenuated by $1/(1 + r^2 |\bm{k}|^2)$, which suppresses features with wavelengths much shorter than $r$ and imposes a minimum feature size of approximately $2\sqrt{3}\, r \approx 17\,\mu$m. Equation~\eqref{eq:filter} is discretized on the same unified grid as the physics, $(r^2 \mathbf{K} + \mathbf{M})\, \boldsymbol{\rho}_{\mathrm{node}} = \mathbf{M} \tilde{\boldsymbol{\rho}}_{\mathrm{node}}$ with $\mathbf{K}$ and $\mathbf{M}$ the assembled stiffness and consistent mass matrices, and element-centered densities are mapped to nodes and back by sparse projection matrices: a node-to-element averager $\mathbf{N}_{2E}$ and a row-normalized element-to-node scatter $\mathbf{E}_{2N}$. The system is solved by Jacobi-preconditioned conjugate gradients wrapped as a custom differentiable operation whose backward pass is a second solve with the same (symmetric) operator.

\FloatBarrier
\subsection{Pairwise Interface-Exclusion Penalty}\label{app:interface} We penalize the adjacency of metallurgically incompatible material pairs (here Ti and Steel), whose direct contact forms brittle Ti--Fe intermetallics \cite{moshokoa2024}. To distinguish sharply between ``material present'' and ``material absent,'' each phase field is first passed through the smooth Heaviside indicator
\begin{equation}\label{eq:heaviside}
\mathcal{H}_{\beta}(\rho_i) = \frac{1}{1 + \exp\!\big[-\beta\, (\rho_i - \eta)\big]},
\end{equation}
with presence threshold $\eta = 0.30$ and sharpness $\beta$. Adjacency is detected by the neighbor average $\bar{\mathcal{H}}_i = \mathbf{N}_{2E} \mathbf{E}_{2N}\, \mathcal{H}_{\beta}(\rho_i)$, which reuses the sparse projections of the Helmholtz filter to average the indicator over the elements incident to each element's nodes. The penalty is the symmetric, volume-normalized overlap
\begin{equation}\label{eq:penalty}
\Pi_{\mathrm{if}} = \frac{1}{|\Omega|} \int_\Omega \Big[ \mathcal{H}_{\beta}(\rho_{\mathrm{Ti}})\, \bar{\mathcal{H}}_{\mathrm{Steel}} + \mathcal{H}_{\beta}(\rho_{\mathrm{Steel}})\, \bar{\mathcal{H}}_{\mathrm{Ti}} \Big]\, \mathrm{d}V,
\end{equation}
evaluated on the filtered field. The symmetry makes the penalty agnostic to which material lies on which side of the interface, and the volume normalization makes its magnitude mesh independent. The continuation schedule reflects a property of Eq.~\eqref{eq:heaviside} on nearly uniform fields: at intermediate densities the indicator floor is not small, so on a grey design the penalty is not interface-selective but acts as a global suppressor of both phases. The penalty is therefore held off ($\omega_{\mathrm{if}} = 0$) while the mass constraint is being ramped and the field binarizes; from the end of the mass ramp, $\omega_{\mathrm{if}}$ is increased linearly from $0$ to $5\times10^{6}$ and the sharpness from $\beta = 4$ to $16$ over the following $2000$ epochs, after which both are held at their maxima.

\FloatBarrier
\subsection{Mass Constraint and Continuation}\label{app:mass}
With the phase mass densities $\bar{\rho}_i$ of Sec.~\ref{sec:tdep}, the physical mass of the filtered design is $m(\boldsymbol{\rho}) = \int_\Omega \sum_i \bar{\rho}_i\, \rho_i(\bm{x})\, \mathrm{d}V$, and the budget is $\psi_m M_0$, where $M_0 = \bar{\rho}_{\mathrm{Cu}} |\Omega|$ is the mass of the domain filled with the densest phase and $\psi_m = 0.25$. The loss carries the squared relative residual, $C_M = m(\boldsymbol{\rho})/(\psi_m M_0) - 1$ in Eq.~\eqref{eq:loss}. The target mass fraction is ramped linearly from its value at the first iteration down to $\psi_m$ over the first half of training, with $C_M^2$ penalized two-sidedly so that the realized mass tracks the schedule; after the ramp the constraint becomes one-sided and penalizes only mass above the budget. 

The SIMP exponent is continuated $p: 1 \to 3$ over the same ramp, and the energy-interpolation indicator $\gamma$ of Eq.~\eqref{eq:wang} is a normalized tanh-Heaviside of the local solid fraction (sharpness $100$, threshold $0.10$), held fixed throughout. Each run comprises $10{,}000$ epochs; the displacement adjoint activates after the 2000-epoch primal warm-up of Sec.~\ref{sec:sens}, and the best design satisfying the stability and adjoint-health criteria of Sec.~\ref{sec:sens} after the mass ramp is retained.

\FloatBarrier
\section{Property-Fit Polynomials and Their Antiderivatives}\label{app:props}
This appendix tabulates the nine polynomial correction factors of Eq.~\eqref{eq:propfactor} and the antiderivatives through which the thermal eigenstrain of Eq.~\eqref{eq:eigenstrain} is evaluated, so that all property inputs of the study can be reproduced exactly. Each factor is a polynomial in $\Delta T = T - 293$~K,
\begin{equation}\label{eq:polyform}
f(\Delta T) = \sum_{k=0}^{n} c_k\, \Delta T^{k}, \qquad c_0 = 1,
\end{equation}
with the coefficients of Table~\ref{tab:propcoef}; the constraint $c_0 = 1$ makes every fit exact at 293 K, where the base values of Sec.~\ref{sec:tdep} apply. The fits are valid on $293\text{--}1100$ K and are held constant outside this window (Sec.~\ref{sec:tdep}). The thermal eigenstrain uses the exact antiderivative of the instantaneous expansion factor,
\begin{equation}\label{eq:antider}
F_{\alpha}(\Delta T) = \int_0^{\Delta T} f_{\alpha}(\tau)\, \mathrm{d}\tau = \sum_{k=0}^{n} \frac{c_k}{k+1}\, \Delta T^{k+1},
\end{equation}
normalized so that $F_\alpha(0) = 0$; its coefficients, obtained term by term from Table~\ref{tab:propcoef}, are listed in Table~\ref{tab:antider}. With $\Tinf = 293$ K, the eigenstrain of Eq.~\eqref{eq:eigenstrain} reduces to $\epsth = \sum_i \rho_i\, \alpha_i F_{\alpha,i}(\Delta T)$, and the secant coefficient of Eq.~\eqref{eq:secant} to $\bar{\alpha}_i(\TD) = \alpha_i F_{\alpha,i}(\Delta T_D)/(\TD - \Tinf)$.
%==============================================
\begin{table*}[h]
\caption{Coefficients $c_k$ (units K$^{-k}$) of the polynomial property factors of Eq.~\eqref{eq:polyform}, fitted over $293\text{--}1100$ K with $c_0 = 1$ exact. Empty entries denote absent orders.}
\label{tab:propcoef}
\centering
\begin{tabular}{llcccc}
\toprule
Factor & Phase & $c_1$ & $c_2$ & $c_3$ & $c_4$ \\
\midrule
\multirow{3}{*}{$f_{\kappa}$}
 & Ti    & $-8.32894\times10^{-4}$ & $\phantom{-}2.02144\times10^{-6}$ & $-1.87281\times10^{-9}$ & $\phantom{-}7.42533\times10^{-13}$ \\
 & Cu    & $-1.72205\times10^{-4}$ & --- & --- & --- \\
 & Steel & $-5.38540\times10^{-4}$ & $-5.82423\times10^{-7}$ & $\phantom{-}5.41279\times10^{-10}$ & --- \\
\addlinespace[2pt]
\multirow{3}{*}{$f_{\alpha}$}
 & Ti    & $\phantom{-}4.50136\times10^{-4}$ & $-1.09250\times10^{-7}$ & --- & --- \\
 & Cu    & $\phantom{-}5.18040\times10^{-4}$ & $-3.84006\times10^{-7}$ & $\phantom{-}4.00361\times10^{-10}$ & --- \\
 & Steel & $\phantom{-}9.45641\times10^{-4}$ & $-2.26320\times10^{-6}$ & $\phantom{-}4.44551\times10^{-9}$ & $-3.06905\times10^{-12}$ \\
\addlinespace[2pt]
\multirow{3}{*}{$f_{E}$}
 & Ti    & $-5.60051\times10^{-4}$ & $\phantom{-}1.04476\times10^{-7}$ & $-2.11509\times10^{-10}$ & --- \\
 & Cu    & $-2.46040\times10^{-4}$ & $-4.35741\times10^{-7}$ & $\phantom{-}2.03154\times10^{-10}$ & --- \\
 & Steel & $-3.20222\times10^{-4}$ & $-1.71611\times10^{-7}$ & $-1.68928\times10^{-10}$ & --- \\
\bottomrule
\end{tabular}
\end{table*}
\begin{table*}[h]
\caption{Coefficients of the expansion antiderivatives $F_{\alpha}(\Delta T) = \sum_k b_k \Delta T^{k}$ of Eq.~\eqref{eq:antider} ($b_k = c_{k-1}/k$, units K$^{-(k-1)}$; $b_1 = 1$ for all phases).}
\label{tab:antider}
\centering
\begin{tabular}{lcccc}
\toprule
Phase & $b_2$ & $b_3$ & $b_4$ & $b_5$ \\
\midrule
Ti    & $\phantom{-}2.25068\times10^{-4}$ & $-3.64167\times10^{-8}$ & --- & --- \\
Cu    & $\phantom{-}2.59020\times10^{-4}$ & $-1.28002\times10^{-7}$ & $\phantom{-}1.00090\times10^{-10}$ & --- \\
Steel & $\phantom{-}4.72821\times10^{-4}$ & $-7.54400\times10^{-7}$ & $\phantom{-}1.11138\times10^{-9}$ & $-6.13810\times10^{-13}$ \\
\bottomrule
\end{tabular}
\end{table*}
%==============================================
\FloatBarrier
\bibliographystyle{unsrt}
\bibliography{refs}

\end{document}